\documentclass[%
 reprint,
 amsmath,amssymb,
 aps,
]{revtex4-1}
\usepackage{multirow}
\usepackage{graphicx}
\usepackage{dcolumn}
\usepackage{bm}
\usepackage{float}
\usepackage{graphicx,subfigure,epsfig}
\usepackage{hyperref}
\hypersetup{
    colorlinks=true,
    citecolor=blue,
    linkcolor=blue,
    filecolor=magenta,
    urlcolor=cyan,}

\begin{document}

\preprint{APS/123-QED}

\title{Echoes of novel black-bounce spacetimes}
\author{Yi Yang}
\email{gs.yangyi17@gzu.edu.cn}
\author{Dong Liu}
 \email{gs.dongliu19@gzu.edu.cn}
\author{Zhaoyi Xu}%
 \email{zyxu@gzu.edu.cn (Corresponding author) }
\author{Yujia Xing}%
\author{Shurui Wu}%
\author{Zheng-Wen Long}%
 \email{zwlong@gzu.edu.cn (Corresponding author) }

\affiliation{College of Physics, Guizhou University, Guiyang, 550025, China}

\date{\today }

\begin{abstract}
Gravitational wave echoes can be used as a significant observable to understand the properties of black holes horizon. In addition, echoes would also closely relate to the unique properties of compact objects. In this work we study the evolution of electromagnetic field and scalar field under the background of novel black-bounce spacetimes. Our results show an obvious echoes signal that can characterize the properties of novel black-bounce spacetimes, and a detailed analysis about the characteristics of the echoes signal is given. By studying the quasinormal ringdown of the three states of novel black-bounce spacetimes, including black holes in $0<a<2 M$, the one-way wormhole in $a=2M$ and the traversable wormhole in $a>2 M$, we find that the echoes signal only appears when $a>2M$ in this spacetime, but when the parameter $a$ increases to a threshold, the echoes signal will be transformed into a quasinormal ringdown of the two-way traversable wormhole.
\end{abstract}

\maketitle


\section{Introduction}
\label{sec:intro}
Black hole physics has recently made very significant progress. In particular, Event Horizon Telescope captured images of the supermassive black hole M87 \cite{M871,M872,M873}, as well as the gravitational waves generated by the merger of compact binaries reported by LIGO Scientific and Virgo Collaboration \cite{ligo1,ligo2,ligo3,ligo4}. By observing the characteristics of gravitational waves generated by the merger of two compact objects, one can use this signal to study the properties of the compact object itself. In addition, it can also give us some new insights about gravitational interaction and astrophysics. Although experimental observations have not detected detailed structures other than the photon sphere, we expect it to give us a deeper understanding of the mechanism of strong gravity in future detections. Particularly, the experimentally accurate detection of quasinormal modes (QNM) may contain some exciting information due to the emergence of some new physics in the late period \cite{cardoso2016}. QNM is a special oscillation that occupies most of the perturbation evolution of the black hole \cite{Kokkotas:1999bd,793,936,163001}. Vishveshwara \cite{936} pointed out that when the background spacetime of a black hole is perturbed, its initial perturbation will gradually be dominated by a damped oscillation with a certain frequency. The frequency and damping characteristics of this oscillation are only related to the spacetime properties of the black hole, and are irrelevant to the initial oscillation. Physicists usually study the physical characteristics of black holes by studying the quasinormal modes of the perturbation \cite{Lijin1,lijin2,lijin3,xip,liudj,csb,csb2}. Reference \cite{moderski3} studied the QNM of dilaton black hole spacetime in scalar field perturbation and Moderski \textit{et al.} also studied the situation of self-interacting scalar field perturbation \cite{moderski2}. The QNM of scalar fields perturbation in the n-dimensional charged black hole spacetime background has been studied in detail in Ref. \cite{moderski1}. The QNM of massive Dirac field perturbaiton in the spherically black hole was studied in Ref. \cite{Moderski:2008nq}.
Reference \cite{cardoso2016} pointed out that extremely compact objects and black holes have a very similar quasinormal ringdown, which makes the detection of gravitational waves not entirely evidence of the existence of black holes, and it is most likely a signal from other compact objects \cite{posada,044027,350,Micchi2020} or wormholes \cite{024016}.

The QNM of wormholes has attracted the attention of many physicists, who have conducted extensive research on it \cite{850,mimi1,mimi2,mimi3,mimi4,mimi5,Liu2020qia,mimi6,mimi7,mimi8,Churilova:2019qph,mimi9,mimi10,mimi11}. Cardoso \textit{et al}. \cite{cardoso2016} discovered the echo signal for the first time in the later period of QNM in wormhole spacetimes. Later, people did further research in a large number of papers on this basis. In Ref. \cite{075014}, they studied the QNM of regular black-hole and wormhole transition and observed unique wormhole echoes picture near the threshold. The metric they considered allows the different parameters corresponding to different spacetime, that is, black holes, the one-way wormhole and the traversable wormhole. Only the traversable wormhole spacetime background has a clear echo picture. In Ref. \cite{084031}, they studied the echoes of several different wormholes with quantum corrections. Bronnikov \textit{et al}. \cite{064004} studied the quasinormal ringdown of the black hole-wormhole transition in brane worlds, and obtained echoes of different wormhole models, the metric of which was proposed by Ref. \cite{024025}.

Recently, Lobo \textit{et al}. proposed a new type of black-bounce spacetimes \cite{metric}, which includes regular black holes and traversable wormholes. In Ref. \cite{franzin}, Franzin \textit{et al}. analyzed the spacetimes geometry of the charged Black-bounce-Reissner-Nordstrom and charged Black-bounce-Kerr-Newman. In Ref. \cite{075014}, the Simpson-Visser black-bounce model proposed by Simpson and Visser \cite{042} is only a special case of novel black-bounce spacetimes. In addition, Mazza \textit{et al}. \cite{Mazza:2021rgq} generalized the Simpson-Visser black bounce metric to the case of rotation. Now we want to study the QNM of this kind of novel black-bounce spacetimes. It is worth noting that we only study a special case in Ref. \cite{metric}, which has the same causal structure as the Simpson-Visser solution, and the Simpson-Visser solution has been analyzed in Ref. \cite{075014}. In this work, studying the ringdown of novel black-bounce spacetimes is our main goal. We will explore the properties of spacetime under different external fields by considering the perturbation of electromagnetic fields and scalar fields. Our results show that there is clear echoes signal in the later stage of the quasinormal ringdown for the novel black-bounce spacetimes.

This paper is organized as follows. In the next section, we introduce the spacetime metric, derive the main equation of spacetime evolution under the influence of electromagnetic field and scalar field perturbation, and obtain the corresponding effective potential. In addition, the image of the effective potential under the perturbation of the electromagnetic field is analyzed. In Sec. \ref{sec3}, we discuss the time domain integration method used in this work. In Sec. \ref{sec4}, we study time-domain profiles of the electromagnetic field and scalar field in the background of the black hole and the wormhole with different spacetime parameters, and analyze in detail different ringdown behaviors including the effects of echo. In Sec. \ref{frequency}, the QNM frequencies of novel black-bounce spacetimes are presented. Finally, we summarize the results of the full text in Sec. \ref{sec5}. In our work, we use geometrized units where $G = c = 1$.
\section{Scalar and electromagnetic field perturbation in novel black-bounce spacetimes}\label{sec2}
We analyze the scalar and electromagnetic perturbation of novel black-bounce spacetimes, which was proposed by Lobo \textit{et al}. Its metric can be written as \cite{metric}
\begin{equation}
d s^{2}=-f(r) d t^{2}+\frac{d r^{2}}{f(r)}+\Sigma^{2}(r)\left(d \theta^{2}+\sin ^{2} \theta d \phi^{2}\right),
\end{equation}
where
\begin{equation}
f(r)=1-\frac{2 M}{\sqrt[4]{r^{4}+a^{4}}}, \Sigma(r)=\sqrt{r^{2}+a^{2}},
\end{equation}
with $M$ being the mass of compact object. Note that for different parameters $a$ corresponds to different spacetimes. When $0<a<2 M$, this spacetime is a regular black hole with two horizons, where regular black hole means that this black hole spacestimes can be extended to $r < 0$. For $a=2 M$, this spacetime is a wormhole with an extremal null throat, which has only extremal horizon. This throat can only travel from one area to another, so it is a one-way traversable wormhole. If $a>2 M$, this spacetime has no horizon and  is a wormhole with a two-way timelike throat. In particular, for $a=0$, this spacetime is Schwarzschild black hole.

In this study, we are interested in the propagation of scalar and electromagnetic fields in above spacetimes. Then we will derive the radial equation of motion and effective potential for scalar field and electromagnetic field perturbation. The general covariant equation of scalar field can be expressed as
\begin{equation}
\frac{1}{\sqrt{-g}} \partial_{\mu}\left(\sqrt{-g} g^{\mu \nu} \partial_{\nu} \Psi\right)=0,
\end{equation}
and the equation of electromagnetic field can be read as
\begin{equation}
\frac{1}{\sqrt{-g}} \partial_{\mu}\left(F_{\rho \sigma} g^{\rho \nu} g^{\sigma \mu} \sqrt{-g}\right)=0,
\end{equation}
where $F_{\rho \sigma}=\partial_{\rho} A^{\sigma}-\partial_{\sigma} A^{\rho}$ and $A_{\mu}$ is a electromagnetic four-potential.

Let us first consider the scalar field. According to the line elements of novel black-bounce spacetimes, we expand the KG equation to arrive
\begin{equation}\label{eq9}
\begin{aligned}
&- \frac{\partial_{t}^{2} \Psi}{f(r)}+\frac{1}{(r^{2}+a^2)}\left(2 r f(r) \partial_{r} \Psi \right. \\
&\left.+(r^{2}+a^2) f^{\prime}(r) \partial_{r} \Psi+(r^{2}+a^2) f(r) \partial_{r}^{2} \Psi\right)\\
&+\frac{1}{(r^{2}+a^2)}\left(\frac{1}{\sin \theta} \partial_{\theta} \sin \theta \partial_{\theta} \Psi+\frac{1}{\sin ^{2} \theta} \partial_{\varphi}^{2} \Psi\right)=0,
\end{aligned}
\end{equation}
where $f^{\prime}(r)$ denote $\frac{d}{d r} f(r)$.
Because of the spherically symmetry of the novel black-bounce spacetimes background, the scalar field $\Psi (t,r, \theta ,\phi)$ can be decomposed into radial and angular parts. Therefore, the scalar field can be assumed to be
\begin{equation}\label{eq6}
\Psi (t,r, \theta ,\phi)=\frac{\psi(t,r)}{R(r)} Y_{l m}(\theta, \varphi),
\end{equation}
where $Y_{l m}(\theta, \varphi)$ represent spherical harmonic function, and $R(r)$ is the function of radial coordinate $r$. Inserting Eq. (\ref{eq6}) into Eq. (\ref{eq9}), we can obtain the following equation
\begin{equation}\label{eq7}
\frac{\partial^{2}\psi(r, t)}{\partial t^{2}}-\frac{\partial^{2}\psi(r, t)}{\partial r_{*}^{2}}+V(r)\psi(r, t) =0,
\end{equation}
where $r_{*}$ is the so-called tortoise coordinate, which is defined as
\begin{equation}
dr_{*}=\frac{1}{f(r)} dr= \frac{1}{1-\frac{2 M}{\sqrt[4]{r^{4}+a^{4}}}}dr.
\end{equation}

\begin{figure*}[t!]\centering
{
\includegraphics[width=0.9\columnwidth]{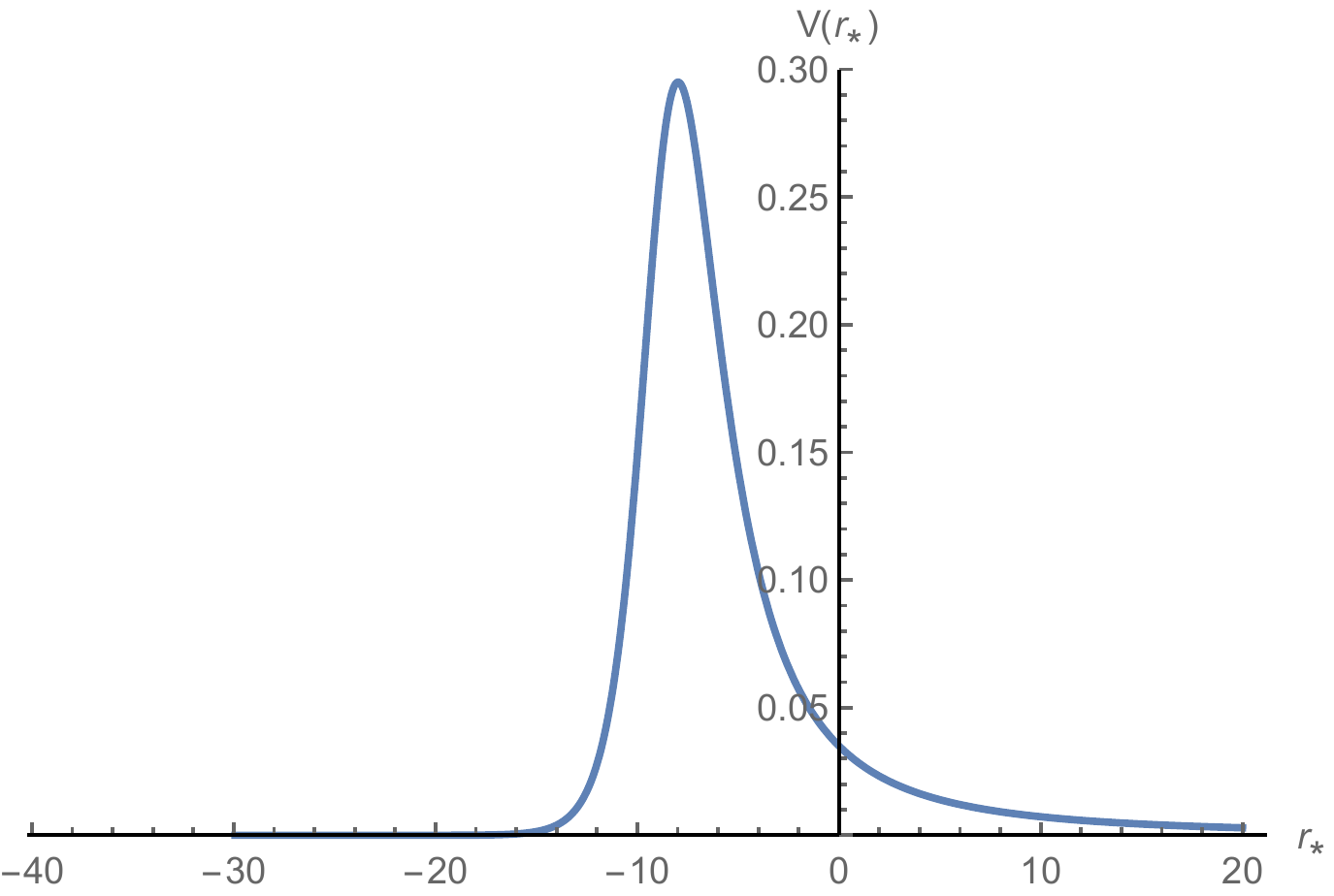}
}
{
\includegraphics[width=0.9\columnwidth]{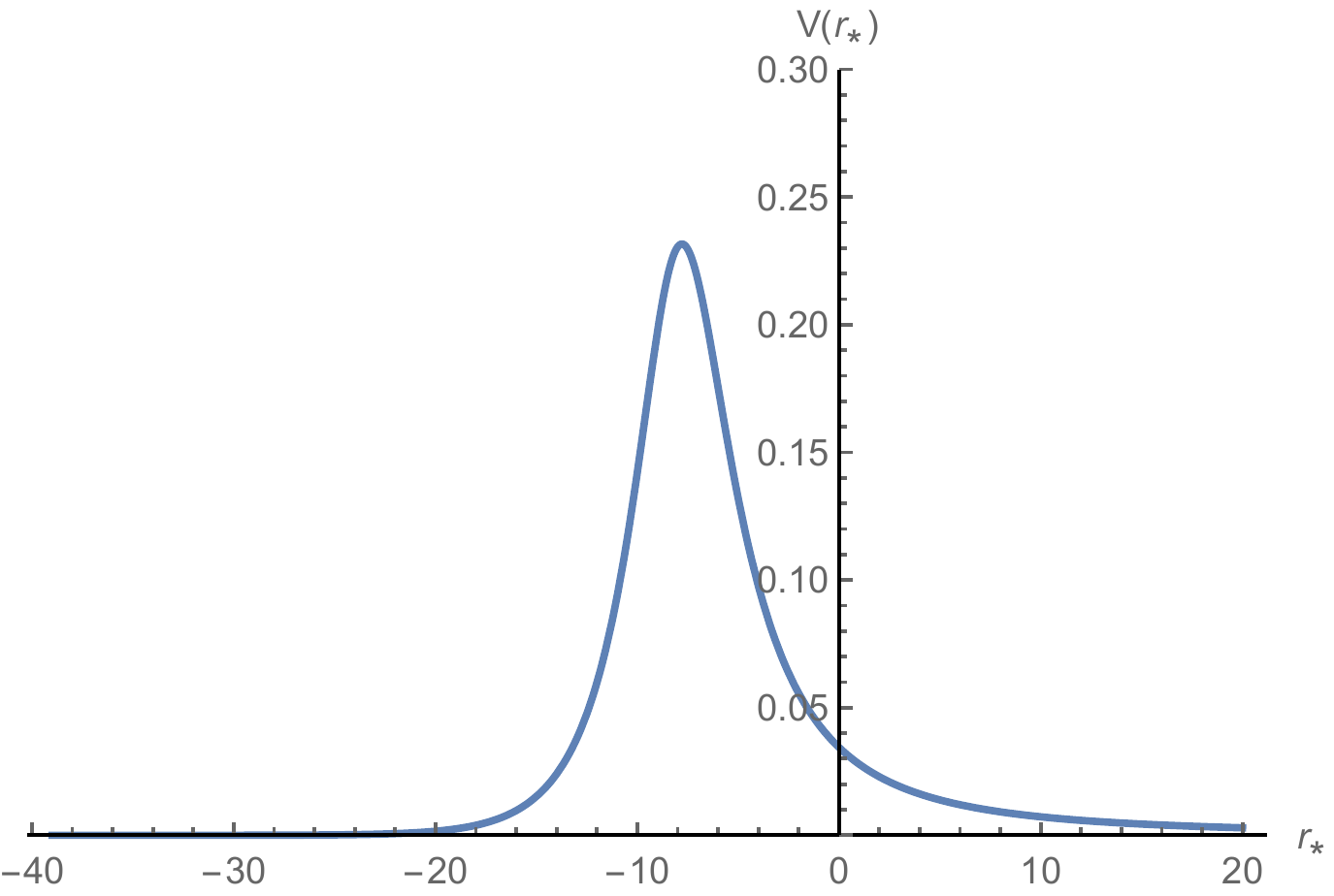}
}
\setlength{\belowcaptionskip}{0.5cm}
\caption{The effective potentials as a function of tortoise coordinate $r_*$ for perturbations of the electromagnetic field on the regular black hole spacetime background with $M=0.5,l=1$, $a=0.1$ (left panel) and $a=0.9$ (right panel).}
\label{V0109}
\end{figure*}

\begin{figure*}[t!]\centering
{
\includegraphics[width=0.9\columnwidth]{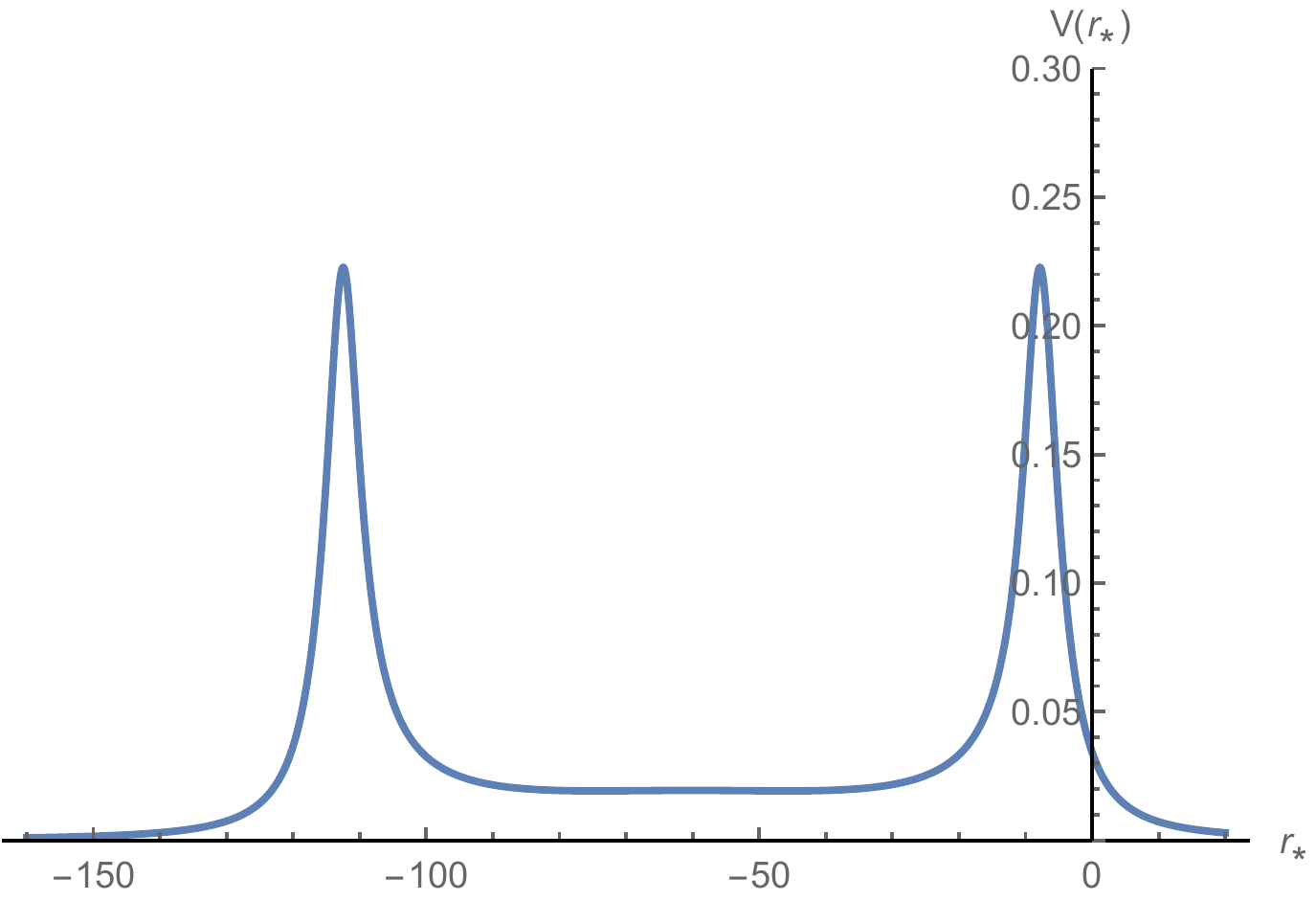}
}
{
\includegraphics[width=0.9\columnwidth]{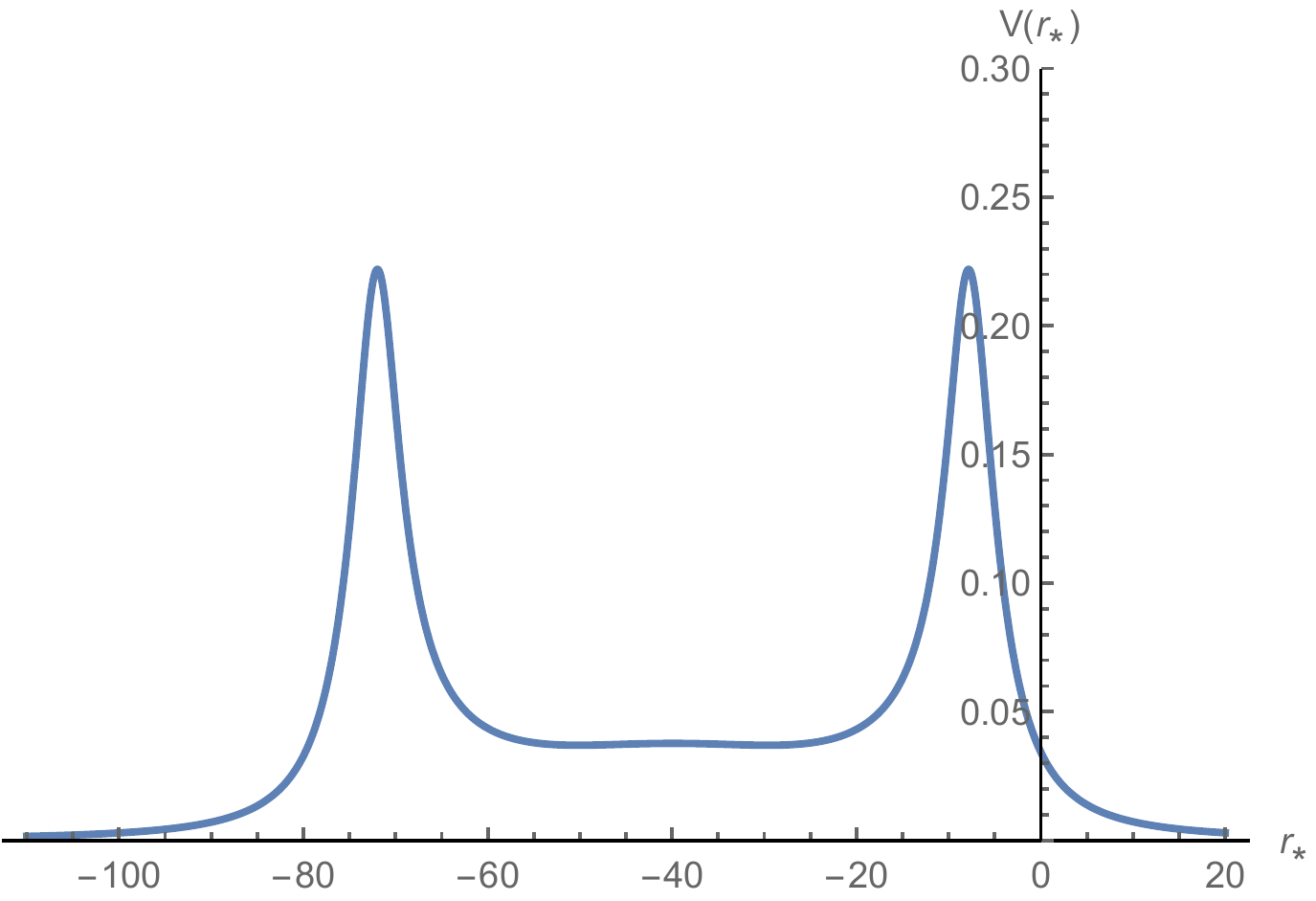}
}
\setlength{\belowcaptionskip}{0.5cm}
\caption{The effective potentials as a function of tortoise coordinate $r_*$ for perturbations of the electromagnetic field with $M=0.5,l=1$, $a=1.01$ (left panel) and $a=1.02$ (right panel).}
\label{V101102}
\end{figure*}

\begin{figure*}[!t]\centering
{
\includegraphics[width=0.9\columnwidth]{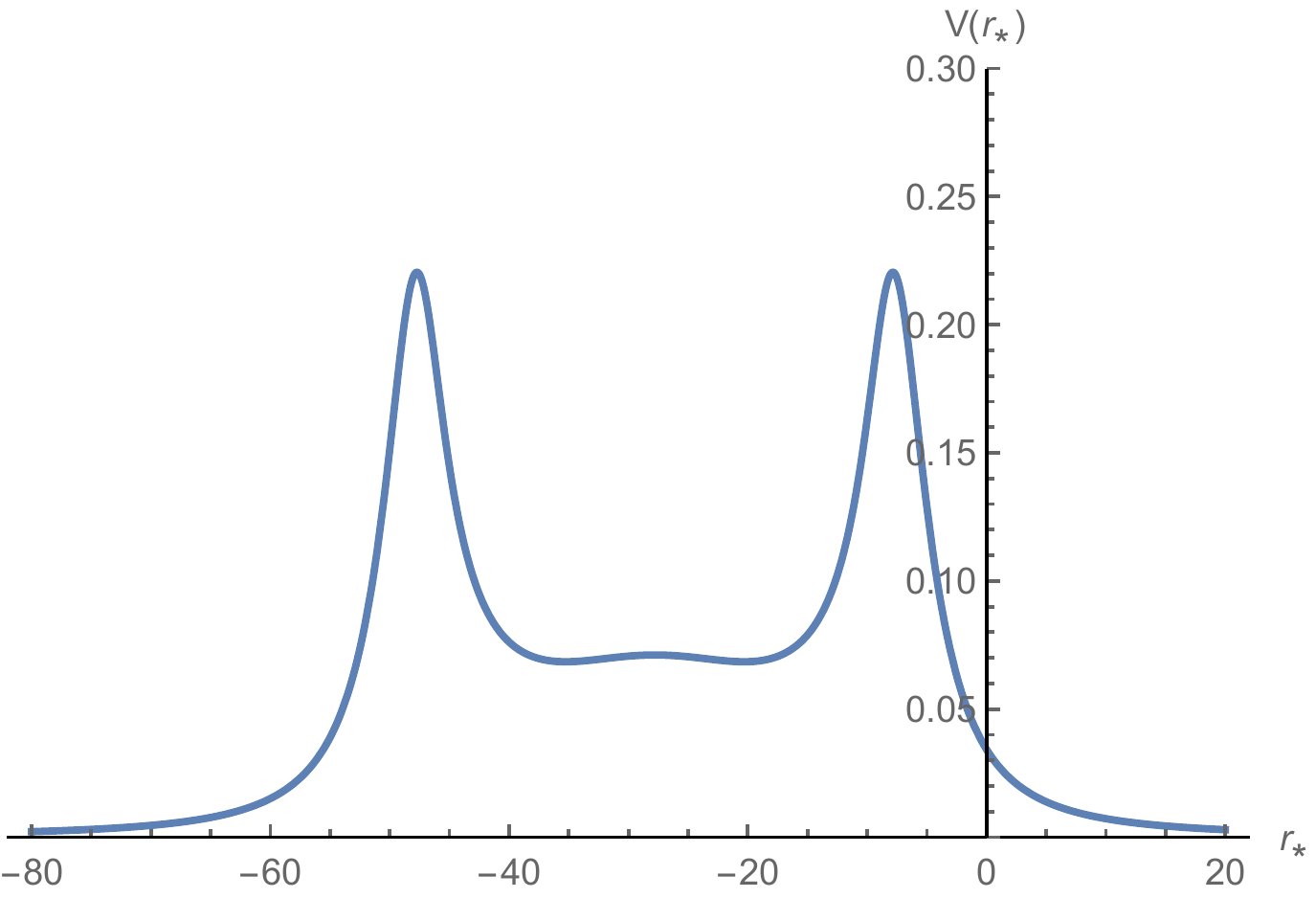}
}
{
\includegraphics[width=0.9\columnwidth]{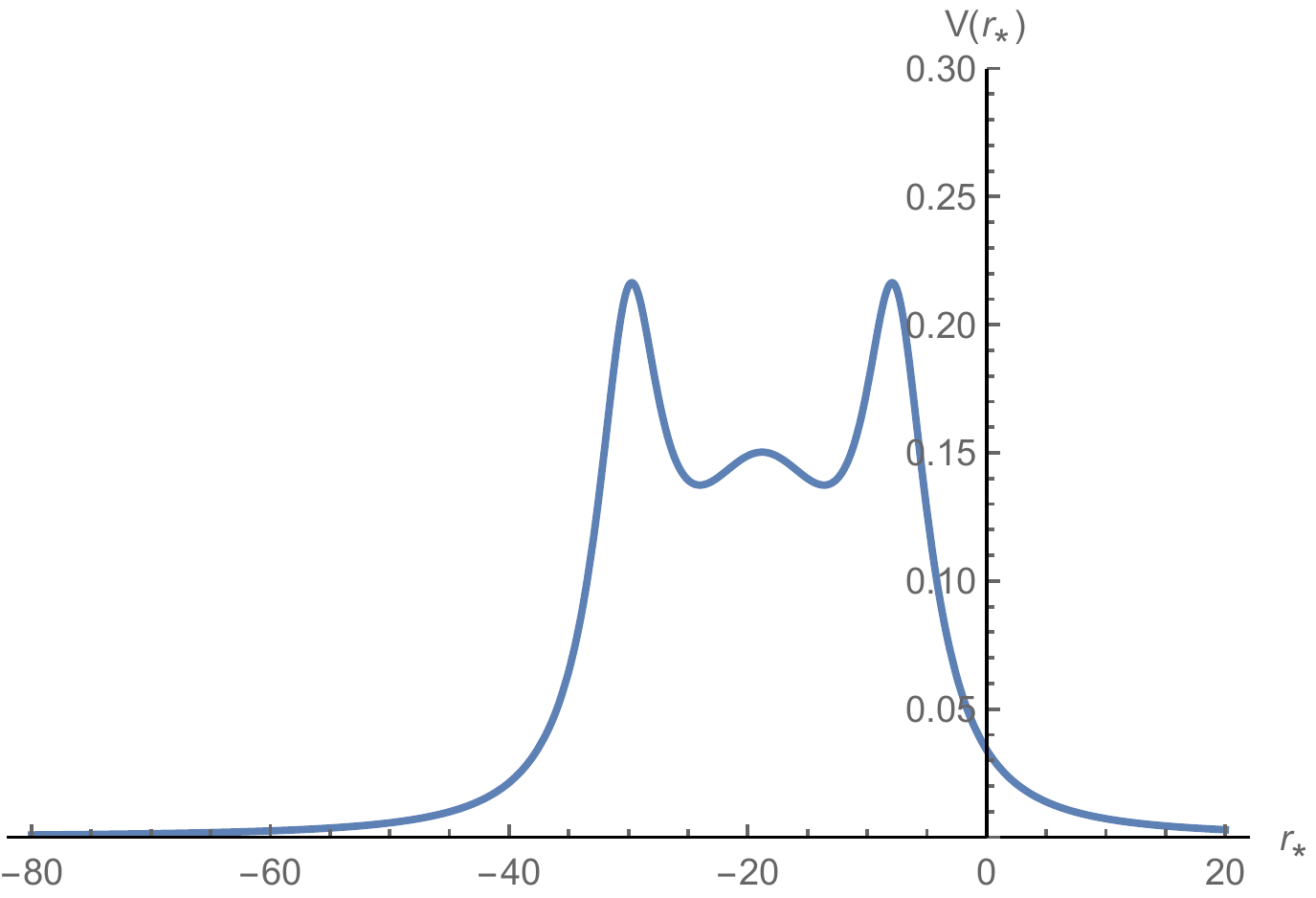}
}
\setlength{\belowcaptionskip}{0.5cm}
\caption{The effective potentials as a function of tortoise coordinate $r_*$ for perturbations of the electromagnetic field with $M=0.5,l=1$, $a=1.04$ (left panel) and $a=1.1$ (right panel).}
\label{V10511}
\end{figure*}

\begin{figure*}[!t]\centering
{
\includegraphics[width=0.9\columnwidth]{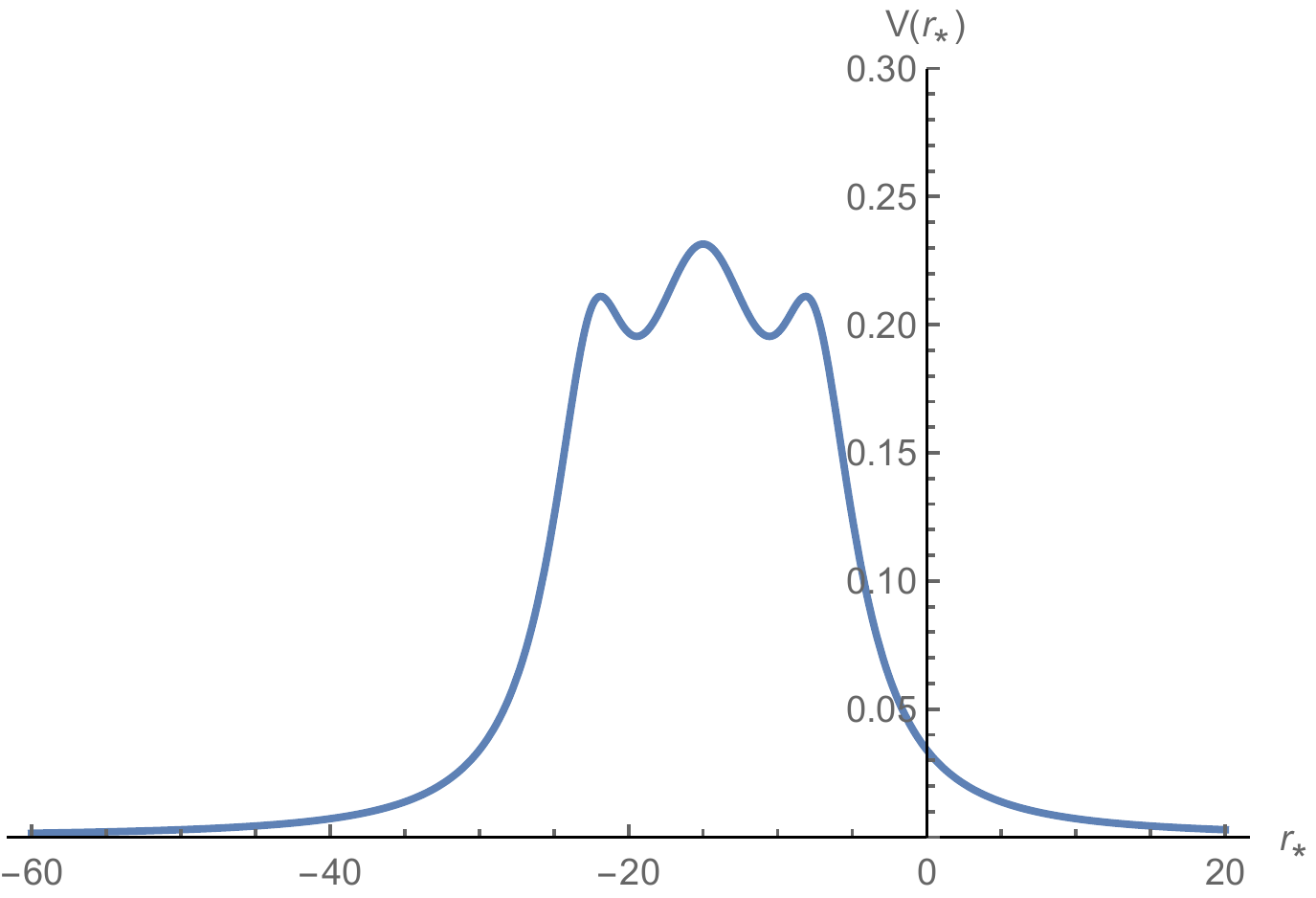}
}
{
\includegraphics[width=0.9\columnwidth]{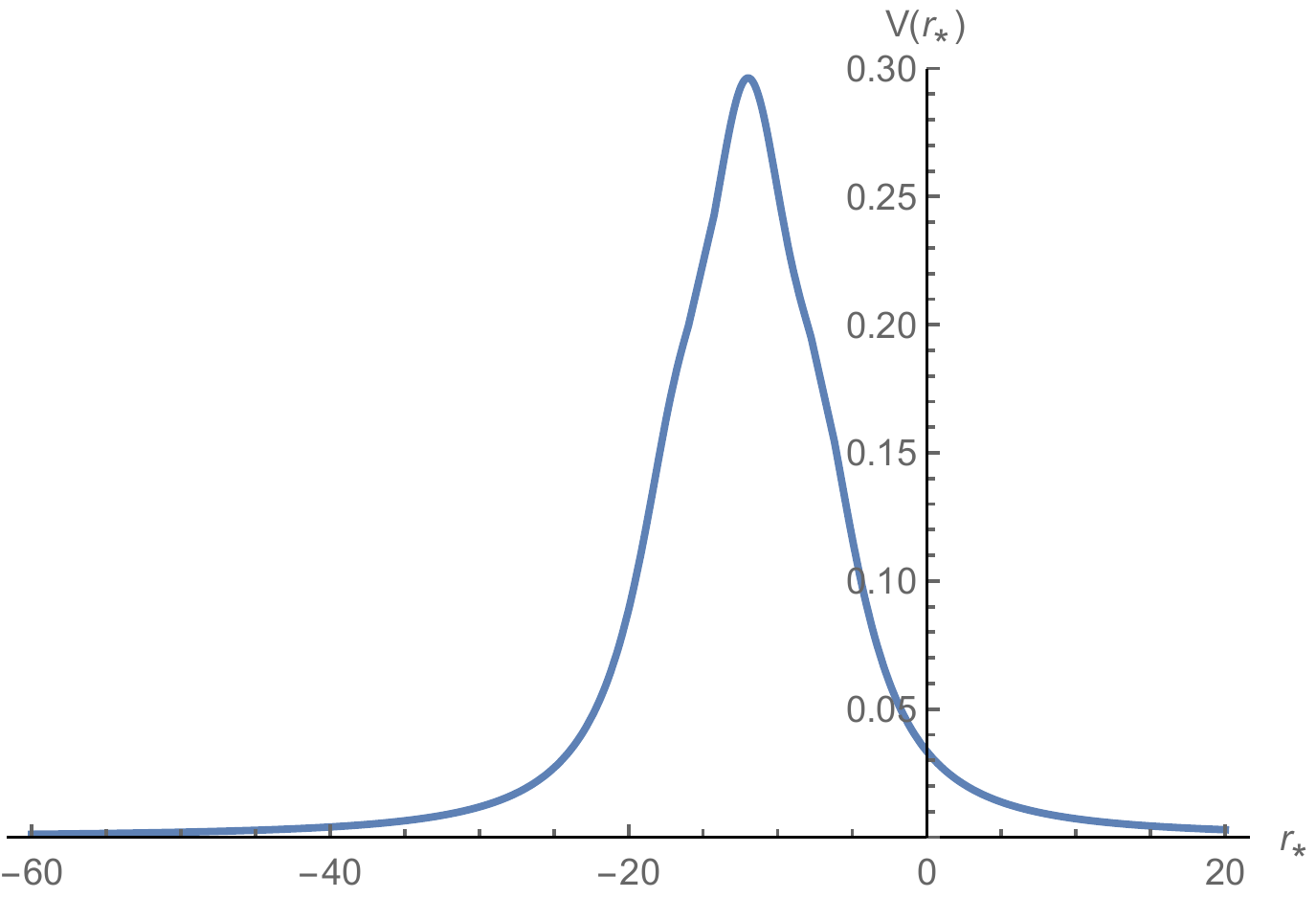}
}
\setlength{\belowcaptionskip}{0.6cm}
\caption{The effective potentials as a function of tortoise coordinate $r_*$ for perturbations of the electromagnetic field with $M=0.5,l=1$, $a=1.2$ (left panel) and $a=1.5$ (right panel).}
\label{V1215}
\end{figure*}

The effective potential $V(r)$ can be written as
\begin{widetext}
\begin{equation}
V(r)=\left(1-\frac{2 M}{\sqrt[4]{r^{4}+a^{4}}}\right)\left(\frac{l(l+1)}{r^{2}+a^{2}}+\frac{2 M r^{6}+a^{2} r^{4}\left(a^{4}+r^{4}\right)^{1 / 4}+a^{6}\left(-2 M+\left(a^{4}+r^{4}\right)^{1 / 4}\right)}{\left(a^{2}+r^{2}\right)^{2}\left(a^{4}+r^{4}\right)^{5 / 4}}\right),
\end{equation}
\end{widetext}
with $l$ being the angular quantum number.

As everyone knows, electromagnetic field perturbation can be divided into odd perturbation and even perturbation, but we find that the effective potentials of odd perturbation and even perturbation are the same. For the perturbation of the electromagnetic field, its equation of motion can also be derived into Eq. (\ref{eq7}), but its effective potential function in the novel black-bounce spacetimes is
\begin{equation}
V(r)=\left(1-\frac{2 M}{\sqrt[4]{r^{4}+a^{4}}}\right)\left(\frac{l(l+1)}{r^{2}+a^2}\right).
\end{equation}

Figures \ref{V0109}-\ref{V1215}  present effective potentials corresponding to different parameters $a$. In Fig. \ref{V0109}, the value of $a$ in the left panel is $0.1$, and in the right panel is $0.9$. In other words, the parameter $a$ is in the interval $(0,2M)$ with $M=0.5$. So Fig. \ref{V0109} is the effective potentials of regular black hole. We can see that regular black hole has only one peak, and there is no divergence behavior. In Figs. \ref{V101102}-\ref{V1215}, the value of parameter $a$ is greater than $2M$, so they are the effective potential of the two-way wormhole. We can see that the effective potential has obvious double peaks, but when the parameter $a$ continues to increase, the effective potential eventually merges to a single peak. It is worth noting that the behavior of merging into one peak is different from the results reported by Ref. \cite{075014}, because we observed three peaks.
\section{Time domain integration method}\label{sec3}
In this section, we will introduce the time domain integration method given by Refs. \cite{price,price2}, which can calculate the evolution of the field over time in a specific spacetime. In order to facilitate the numerical calculation of time domain integration, the light cone coordinates are introduced
\begin{equation}
\begin{array}{l}
u=t-r_*, \\
v=t+r_*,
\end{array}
\end{equation}
where $u$ and $v$ are integral constants. When $r>2GM$ ($G$ denotes the gravitational constant), $u$ describes the radial inward movement of light, and $v$ describes the radial outward movement of light. Using the light cone coordinates, the Eq. (\ref{eq7}) can be expressed as
\begin{equation}\label{eq15}
\frac{\partial^{2}}{\partial u \partial v} \psi(u, v)=-\frac{1}{4}V(r) \psi(u, v).
\end{equation}
The time evolution operator can be expressed as
\begin{equation}
\begin{array}{l}
\exp \left(h \frac{\partial}{\partial t}\right)=\exp \left(h \frac{\partial}{\partial u}+h \frac{\partial}{\partial v}\right)= \\[3pt]
=\exp \left(h \frac{\partial}{\partial u}\right)+\exp \left(h \frac{\partial}{\partial v}\right)-1+ \\[3pt]
+\frac{h^{2}}{2}\left(\exp \left(h \frac{\partial}{\partial u}\right)+\exp \left(h \frac{\partial}{\partial v}\right)\right) \frac{\partial^{2}}{\partial u \partial v}+\mathcal{O}\left(h^{4}\right) .
\end{array}
\end{equation}
Using this operator and Taylor's theorem, the Eq. (\ref{eq15}) can be written discretely as
\begin{equation}
\psi_{N}=\psi_{E}+\psi_{W}-\psi_{S}-\delta u \delta v V\left(\frac{\psi_{W}+\psi_{E}}{8}\right)+O\left(\Delta^{4}\right).
\end{equation}
The points $S, W, E, N$ are defined as follows: $S=(u,v),W=(u+\delta u,v),E=(u,v+\delta v),N=(u+\delta u,v+\delta v)$.
We discover the fact that the damping and oscillation process of scalar field and electromagnetic field perturbation are not sensitive to initial conditions. We use a Gaussian pulse with a width $\sigma$ at the point $(u_0, v_0)$ as the initial pulse. The center of this Gaussian pulse is at $v_c$ and the field is set to zero at $u=u_0$ and $v=v_0$, namely
\begin{equation}
\begin{array}{l}
\psi\left(u=u_{0}, v\right)=\exp \left[-\frac{\left(v-v_{c}\right)^{2}}{2 \sigma^{2}}\right], \\
\psi\left(u, v=v_{0}\right)=0.
\end{array}
\end{equation}
After discretizing the equation and setting the initial conditions, the four-point difference method can be used for numerical calculation. On the $u-v$ plane, we can always calculate the value of the fourth point from the previous three known points so that the calculation can be continued until the value of all points on the $u-v$ plane is calculated. As long as our grid is large enough, we can obtain a good approximate solution to the wave equation.
\begin{figure*}[!t]\centering
{
\includegraphics[width=0.9\columnwidth]{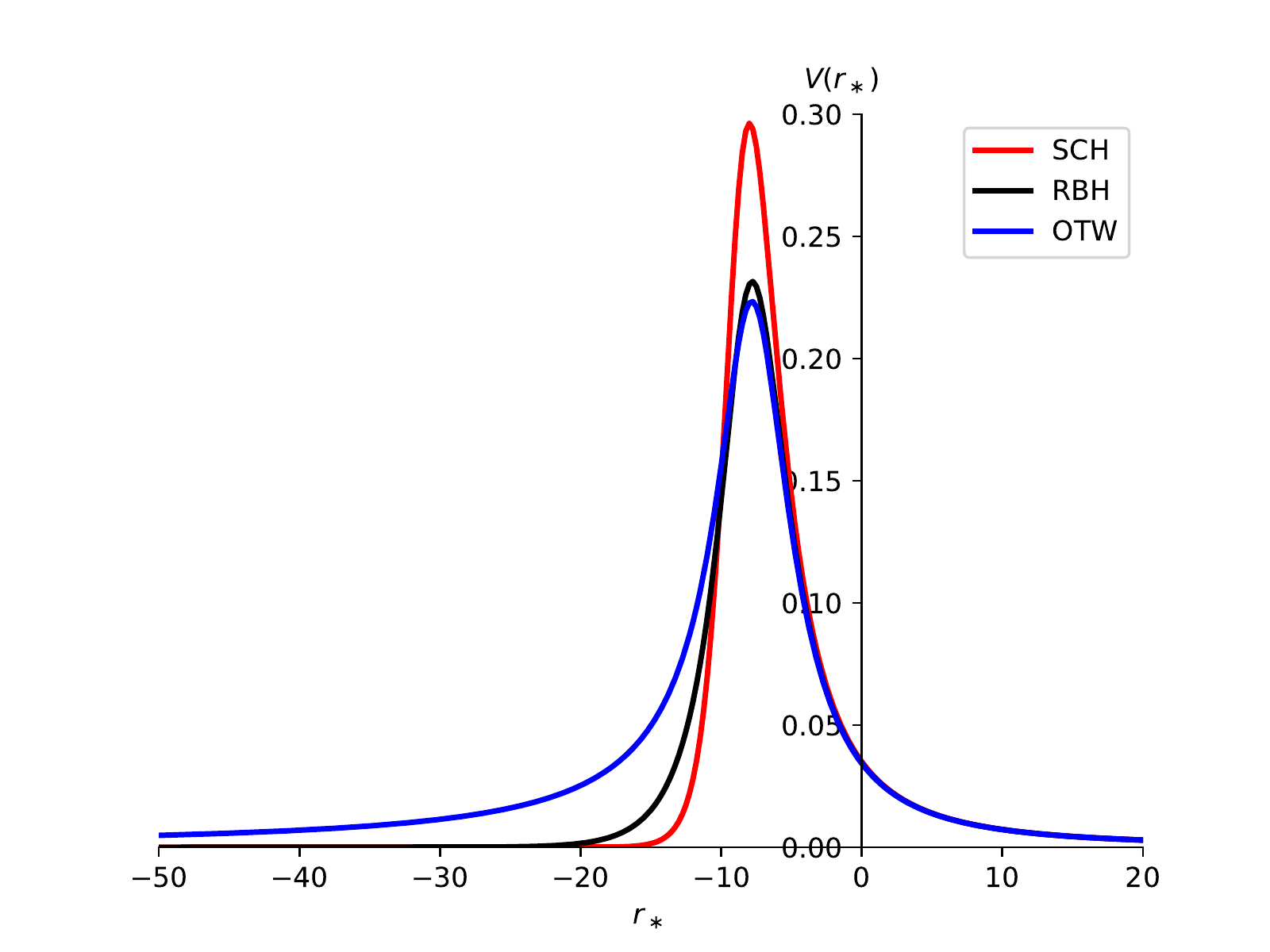}
}
{
\includegraphics[width=0.9\columnwidth]{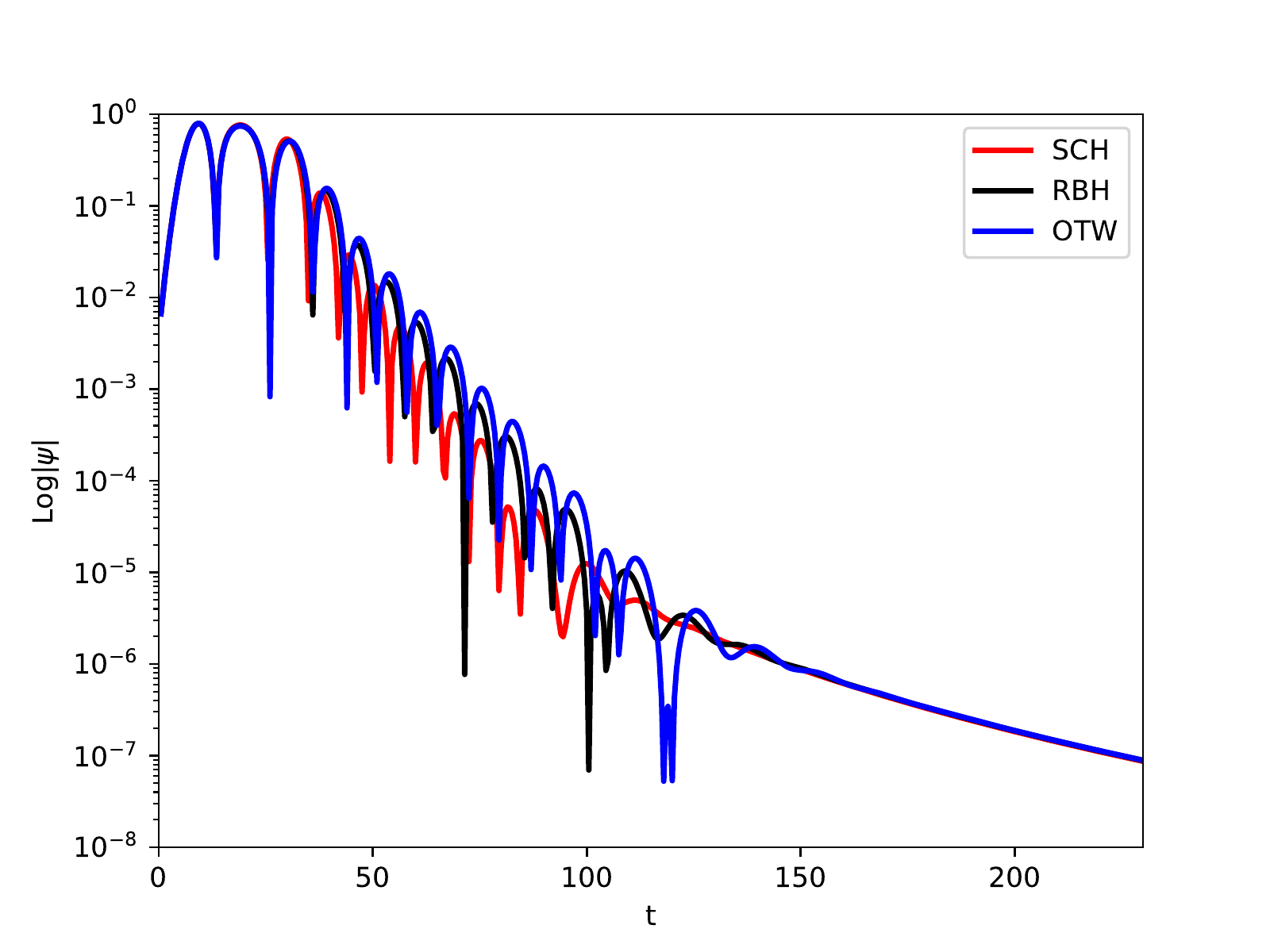}
}
\setlength{\belowcaptionskip}{0.5cm}
\caption{The effective potentials (left panel) and quasinormal ringdown of electromagnetic perturbations (right panel) with $M=0.5,l=1$, for different $a$ ($a=0$ for the Schwarzschild black hole (SCH), $a=0.9$ for the regular black hole (RBH), and $a=1$ for the one-way traversable wormhole (OTW)).}
\label{sanv}
\end{figure*}
\begin{figure*}[!t]\centering
{
\includegraphics[width=0.9\columnwidth]{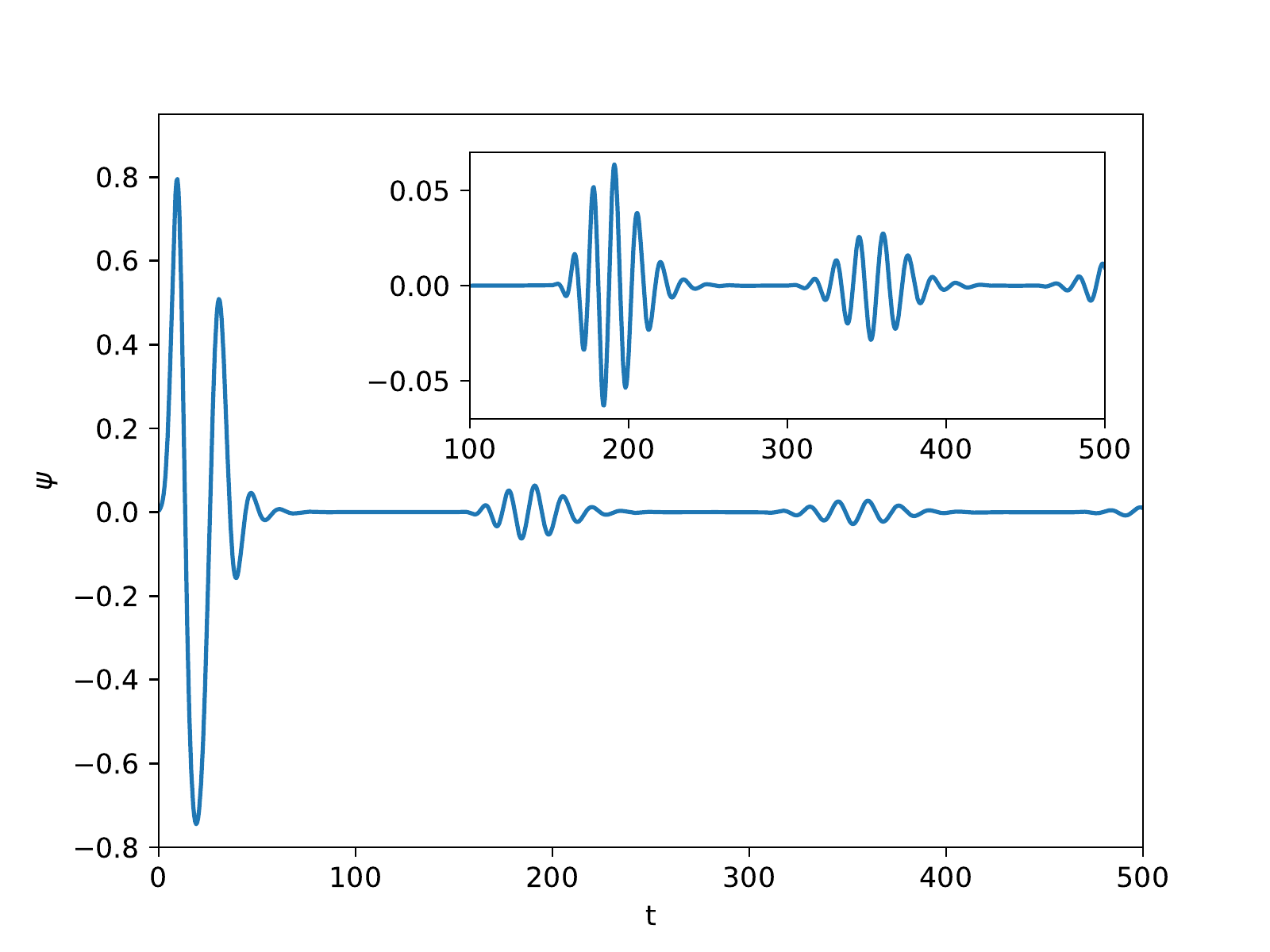}
}
{
\includegraphics[width=0.9\columnwidth]{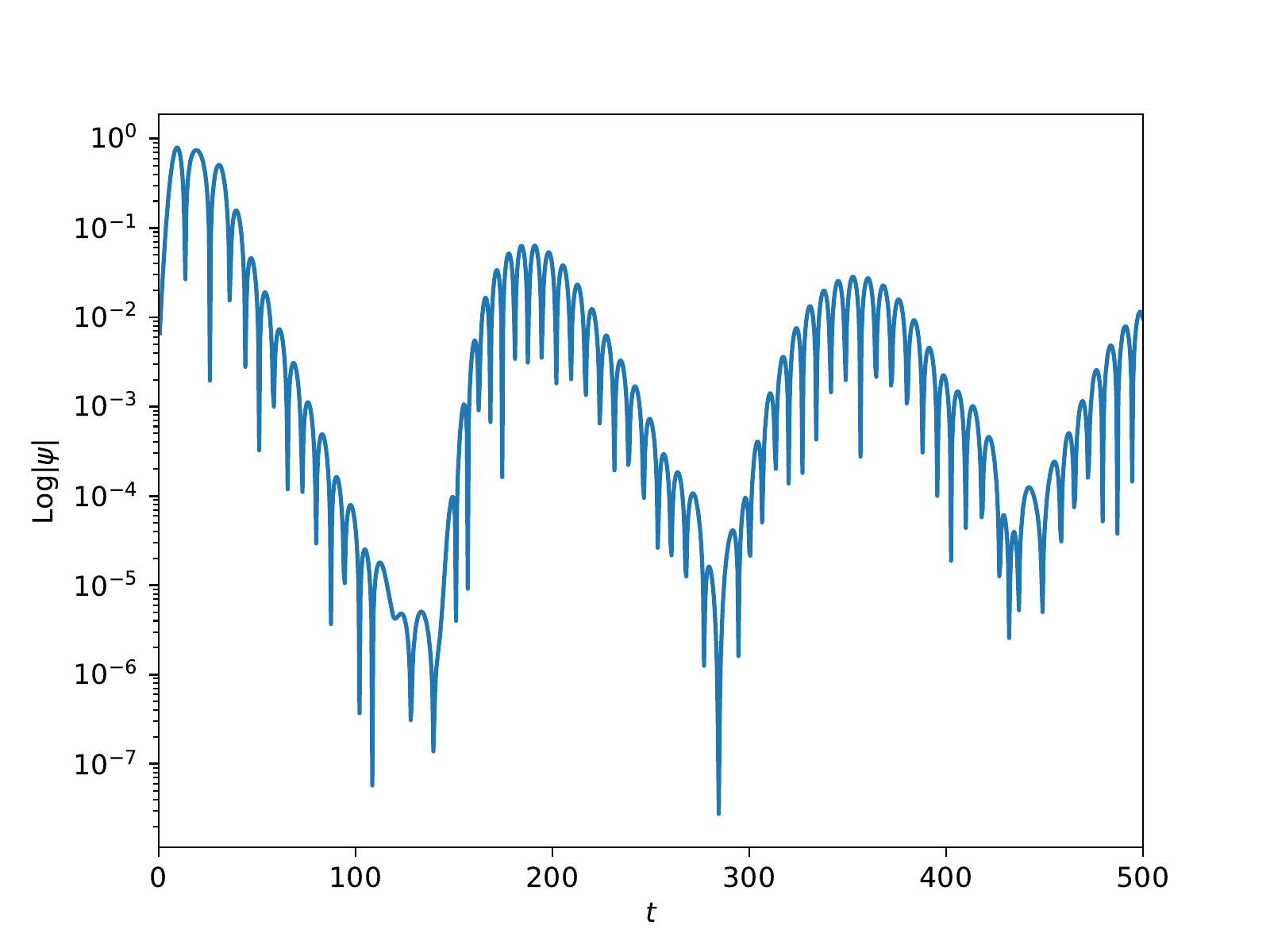}
}
\setlength{\belowcaptionskip}{0.5cm}
\caption{Time evolution of electromagnetic perturbations (left panel) and the semi-logarithmic plot of evolution (right panel) with $M=0.5,l=1,a=1.02$ for the two-way traversable wormhole.}
\label{echo102}
\end{figure*}
\section{The pictures of echoes for novel black-bounce spacetimes}\label{sec4}
In this section, we focus on the echoes signals from electromagnetic field and scalar field perturbations. Using the above numerical calculation process, we study the time evolution under the perturbations of the massless scalar field and electromagnetic field in the novel black-bounce spacetimes background. Note that the necessary condition for the appearance of the echoes is that the potential well must appear. Through the analysis of the potential function in Sec. \ref{sec2}, we find that potential well appears when $a>2M$. Therefore, we will fix the angular quantum number $l=1$ to explore the influence of the spacetime parameter $a$ on echoes when the parameter $a$ is greater than $2M$. In the following discussion, we take $M=0.5$ (unless otherwise specified).

\begin{figure*}[!t]\centering
{
\includegraphics[width=0.9\columnwidth]{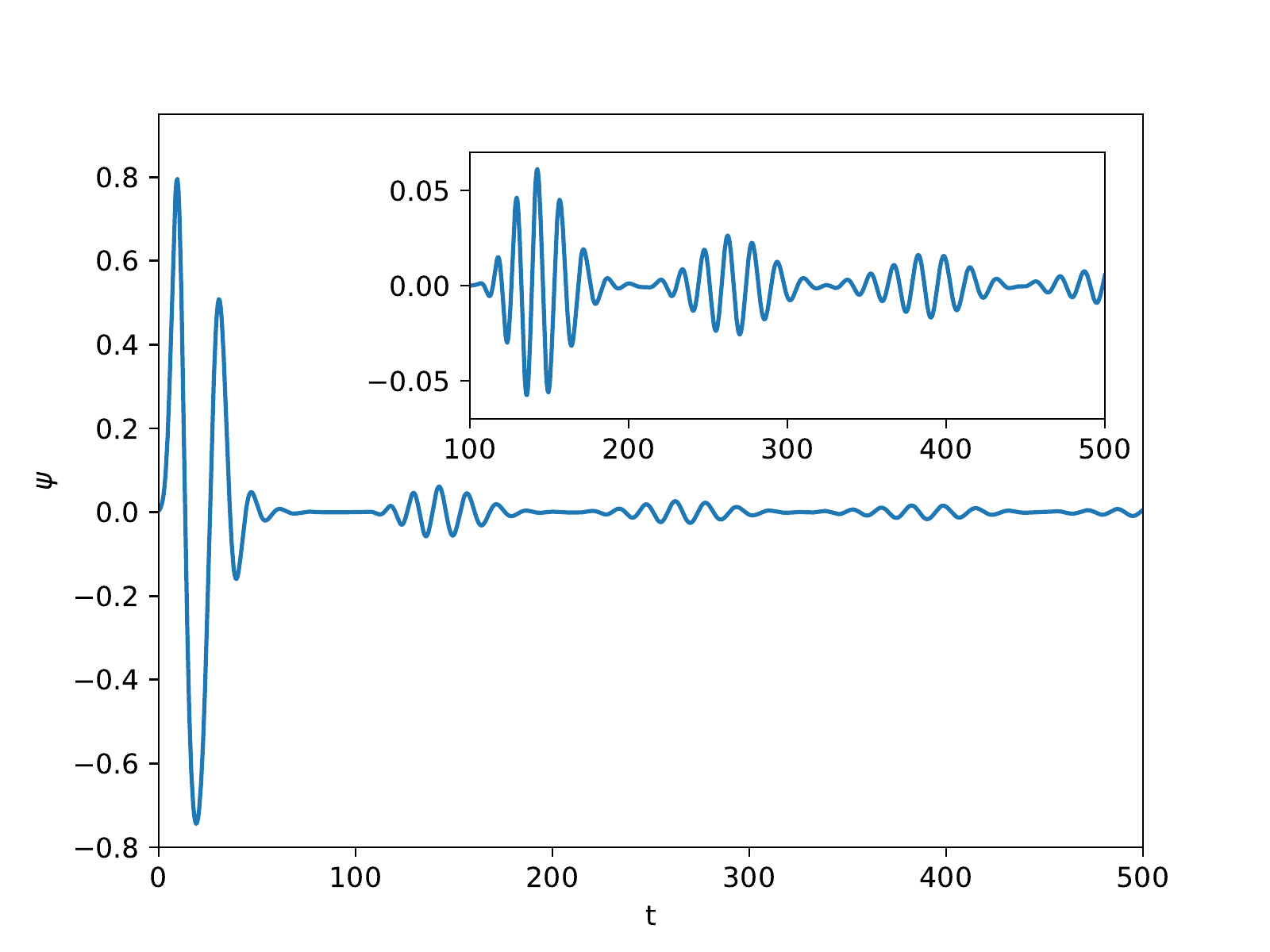}
}
{
\includegraphics[width=0.9\columnwidth]{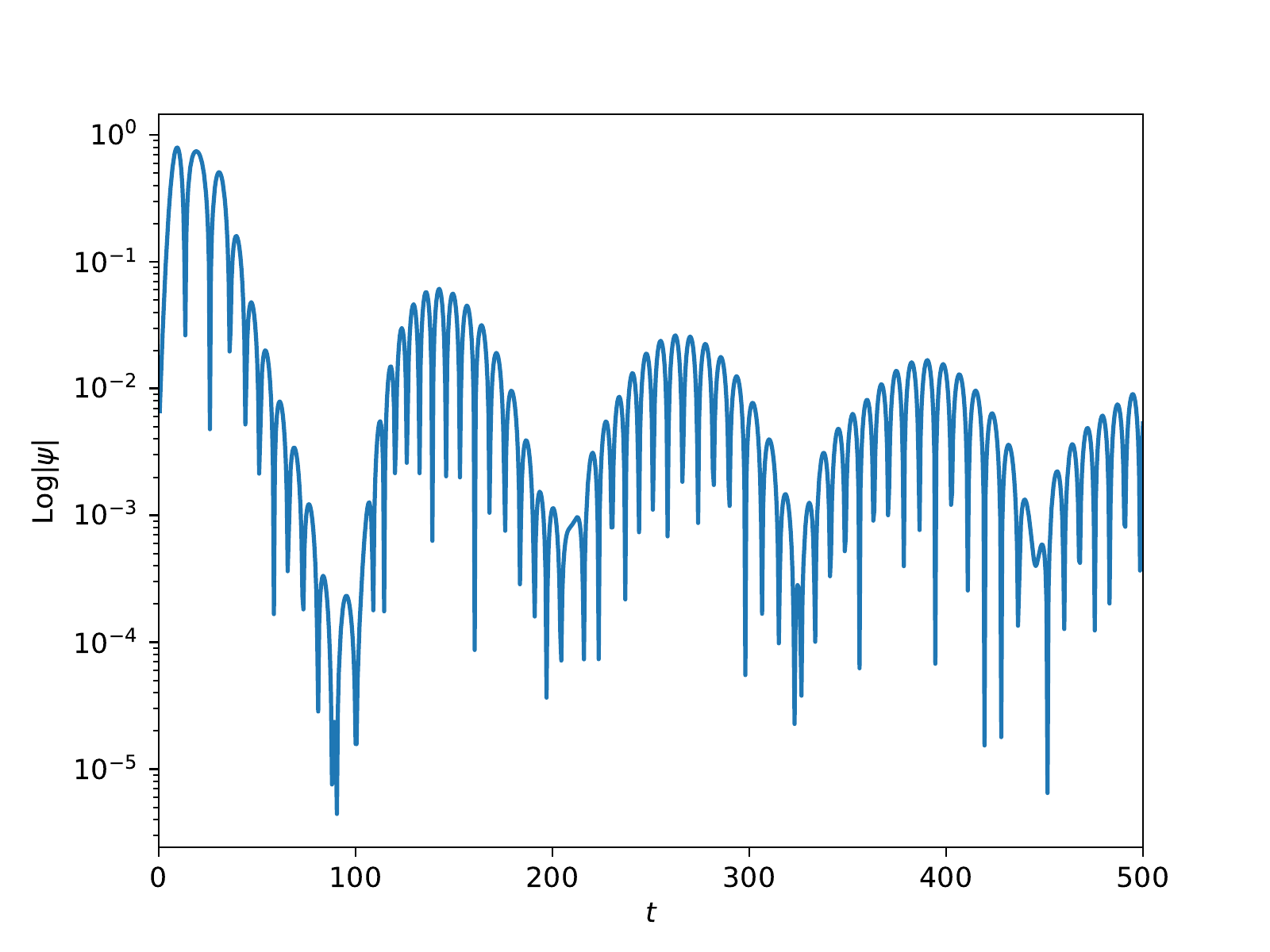}
}
\setlength{\belowcaptionskip}{0.6cm}
\caption{Time evolution of electromagnetic perturbations (left panel) and the semi-logarithmic plot of evolution (right panel) with $M=0.5,l=1,a=1.04$ for the two-way traversable wormhole.}
\label{echo104}
\end{figure*}

\begin{figure*}[!t]\centering
{
\includegraphics[width=0.9\columnwidth]{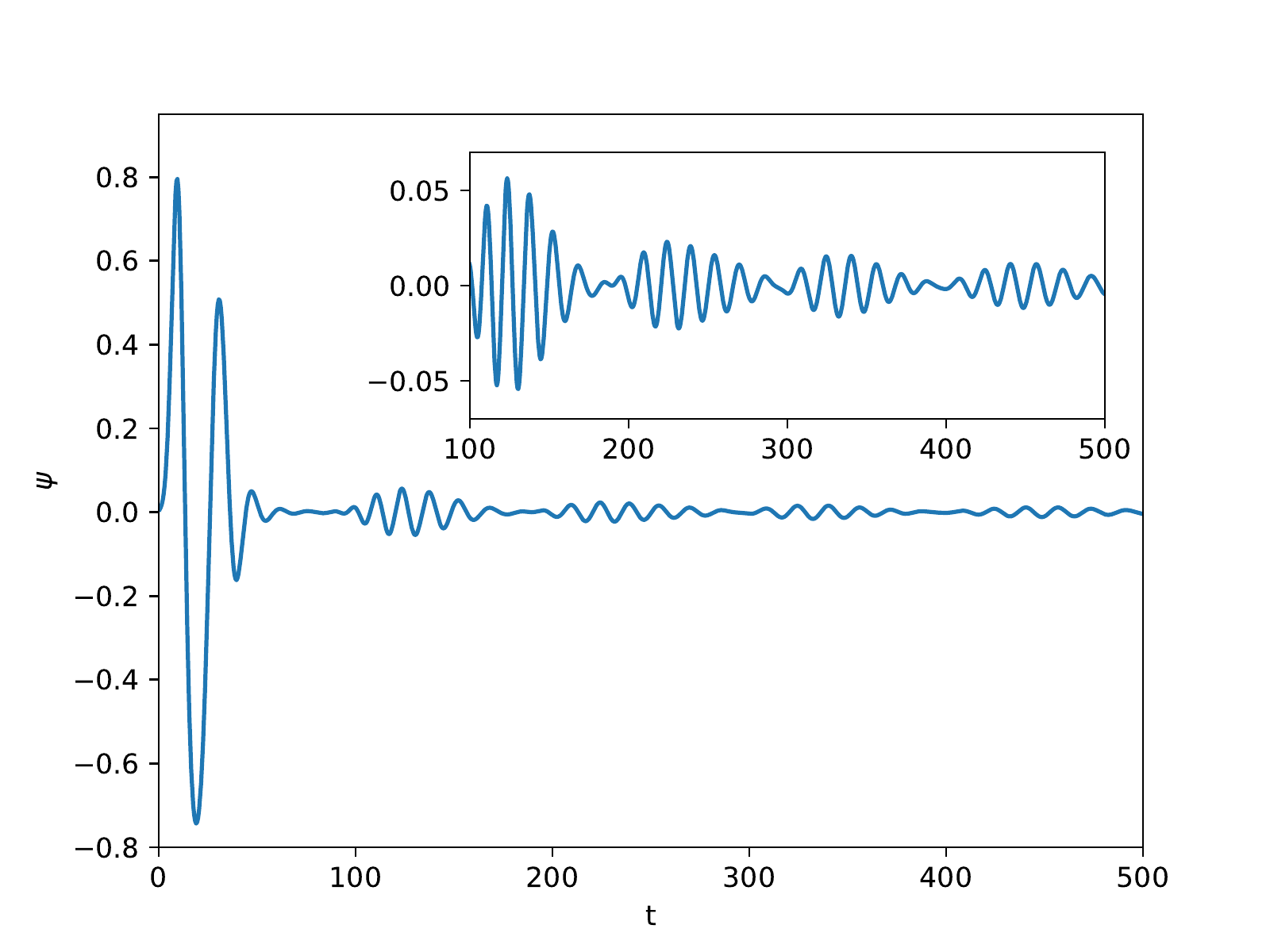}
}
{
\includegraphics[width=0.9\columnwidth]{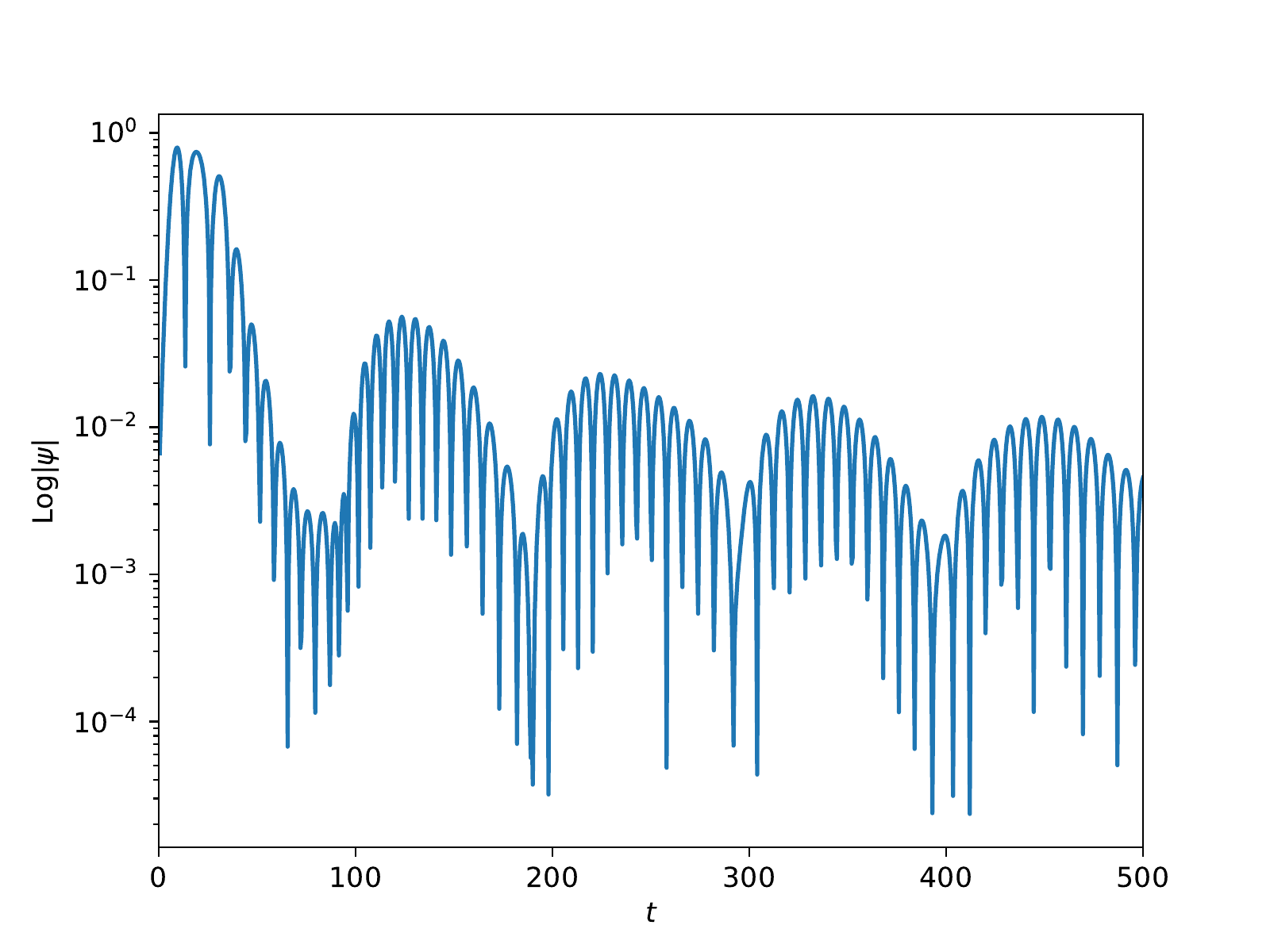}
}
\setlength{\belowcaptionskip}{0.6cm}
\caption{Time evolution of electromagnetic perturbations (left panel) and the semi-logarithmic plot (right panel) of evolution with $M=0.5,l=1,a=1.06$ for the two-way traversable wormhole.}
\label{echo106}
\end{figure*}

For the case of $0\leq a\leq2M$, although there is no potential well in the potential function, we still need to verify it. Therefore, before studying the echo, let's discuss the case of $a\leq2M$. In Fig. \ref{sanv}, we show the effective potentials (left panel) and quasinormal ringdown of electromagnetic perturbations (right panel), where the red line corresponds to the Schwarzschild black hole ($a=0$), the black line corresponds to regular black hole ($a=0.9$), and the blue line corresponds to the one-way traversable wormhole ($a=1$). Regardless of whether it is a regular black hole or a one-way wormhole, there is no double peak in their effective potential. As expected, in quasinormal ringdown figures there is no echo signal, but we can clearly see the three stages. The first is the initial wave burst stage, which is mainly related to the initial perturbation source; the second stage is the QNM stage, which has nothing to do with the initial disturbance and mainly reflects the property of spacetime; the third stage is the tailing stage. Moreover, we found that compared with the quasinormal ringdown of the Schwarzschild black hole, the quasinormal ringdown of a one-way traversable wormhole has the slowest decay rate, followed by regular black holes.

Figures \ref{echo102}-\ref{echo106} show electromagnetic perturbations of novel black-bounce spacetimes for the case of $a>2M$. In these figures, the values of $a$ are very close to $2M$, but they are still greater than $2M$. From these figures, we can see that the clear and unique echoes signal appears after the initial ringdown. These echoes pictures present the following characteristics:

\begin{figure*}[!htpb]\centering
{
\includegraphics[width=0.9\columnwidth]{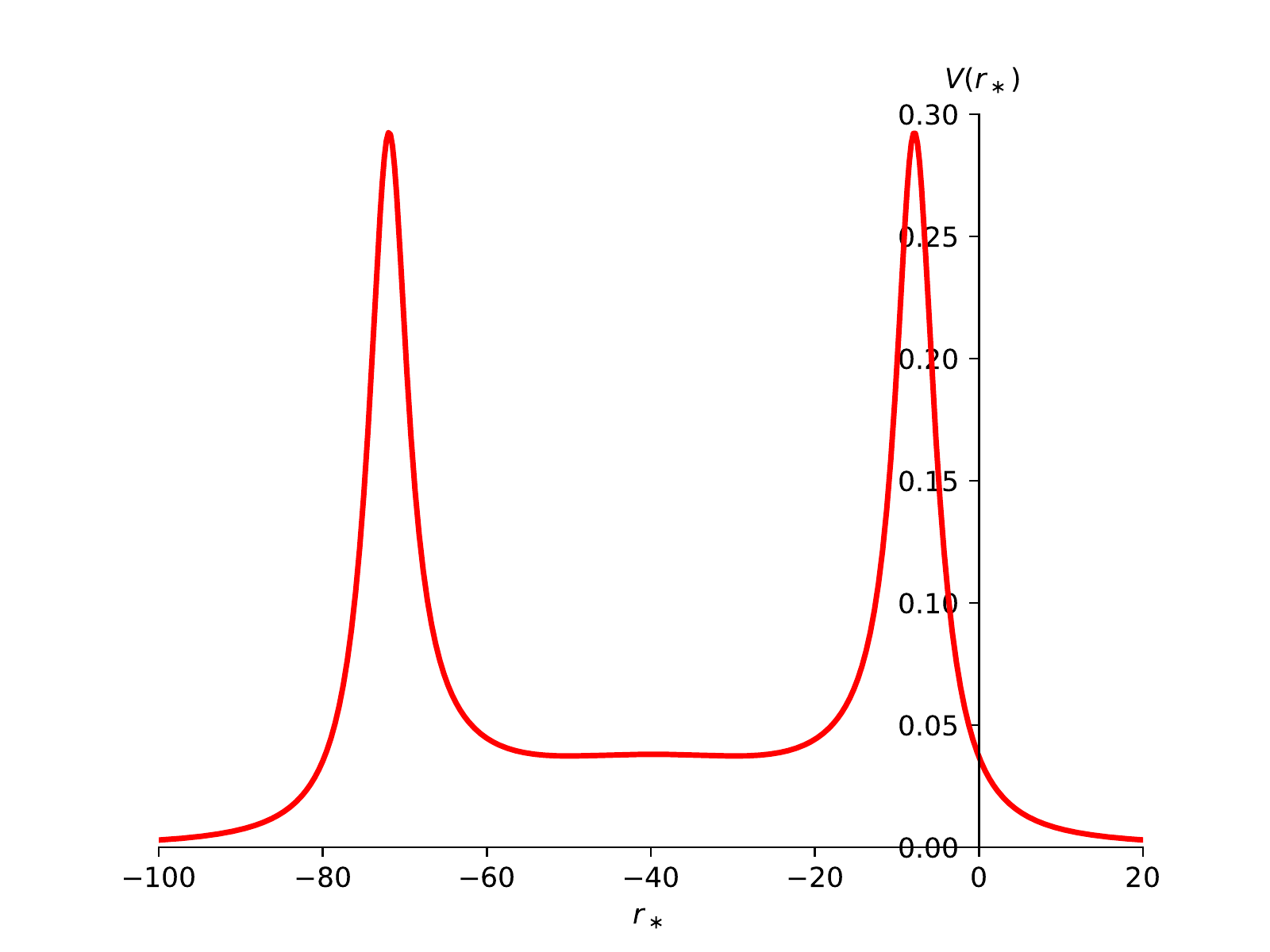}
}
{
\includegraphics[width=0.9\columnwidth]{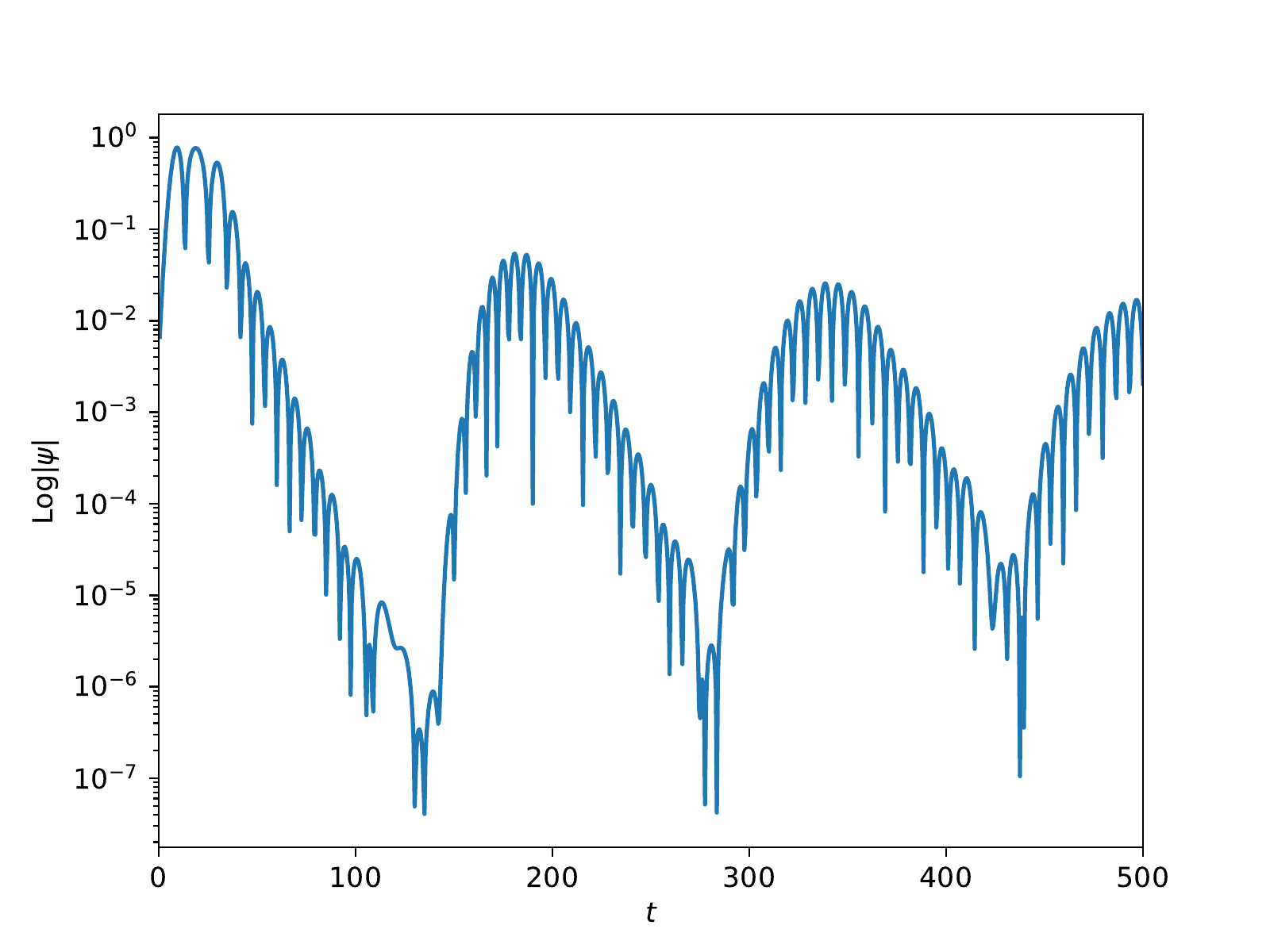}
}
{
\includegraphics[width=0.9\columnwidth]{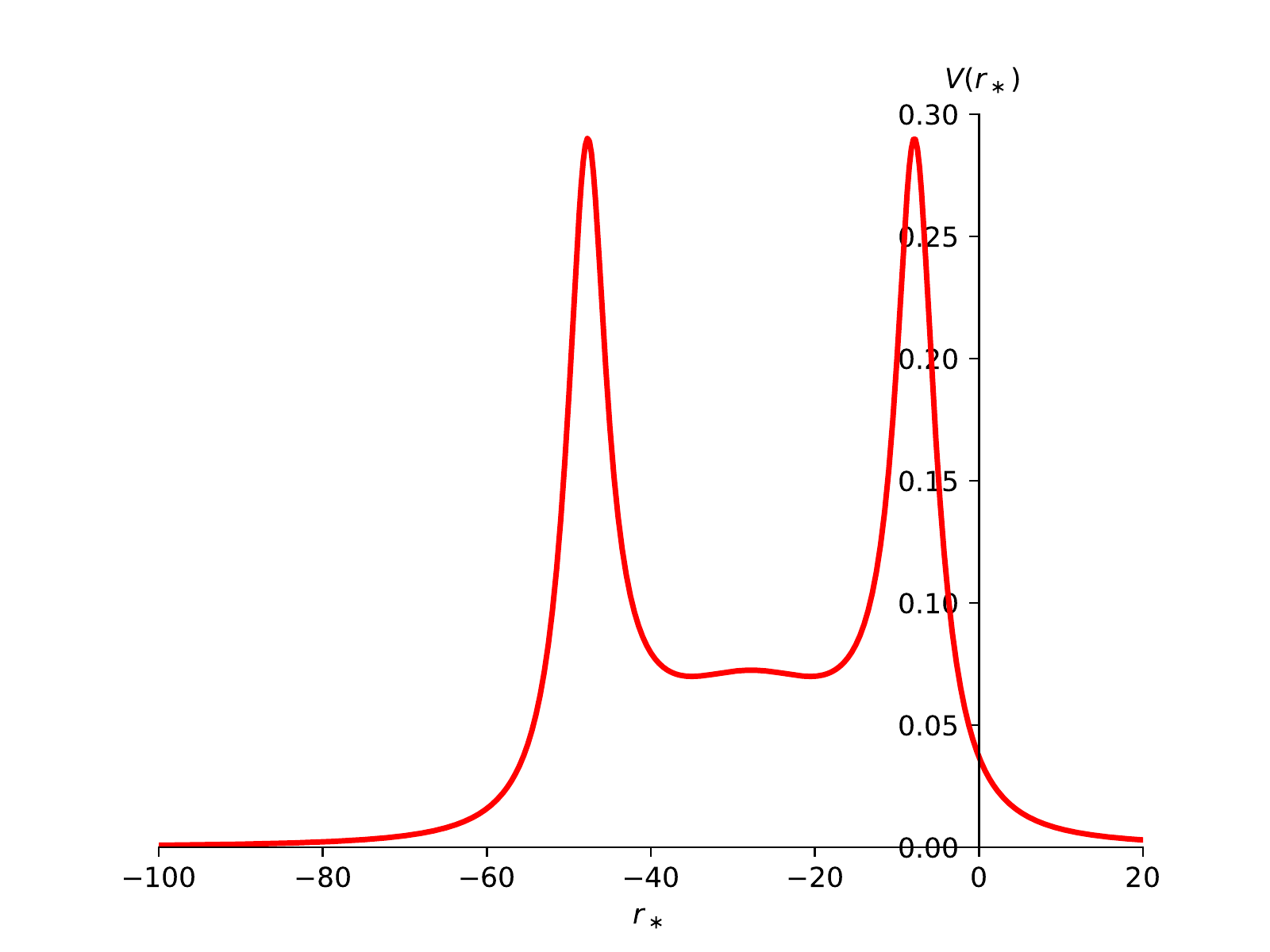}
}
{
\includegraphics[width=0.9\columnwidth]{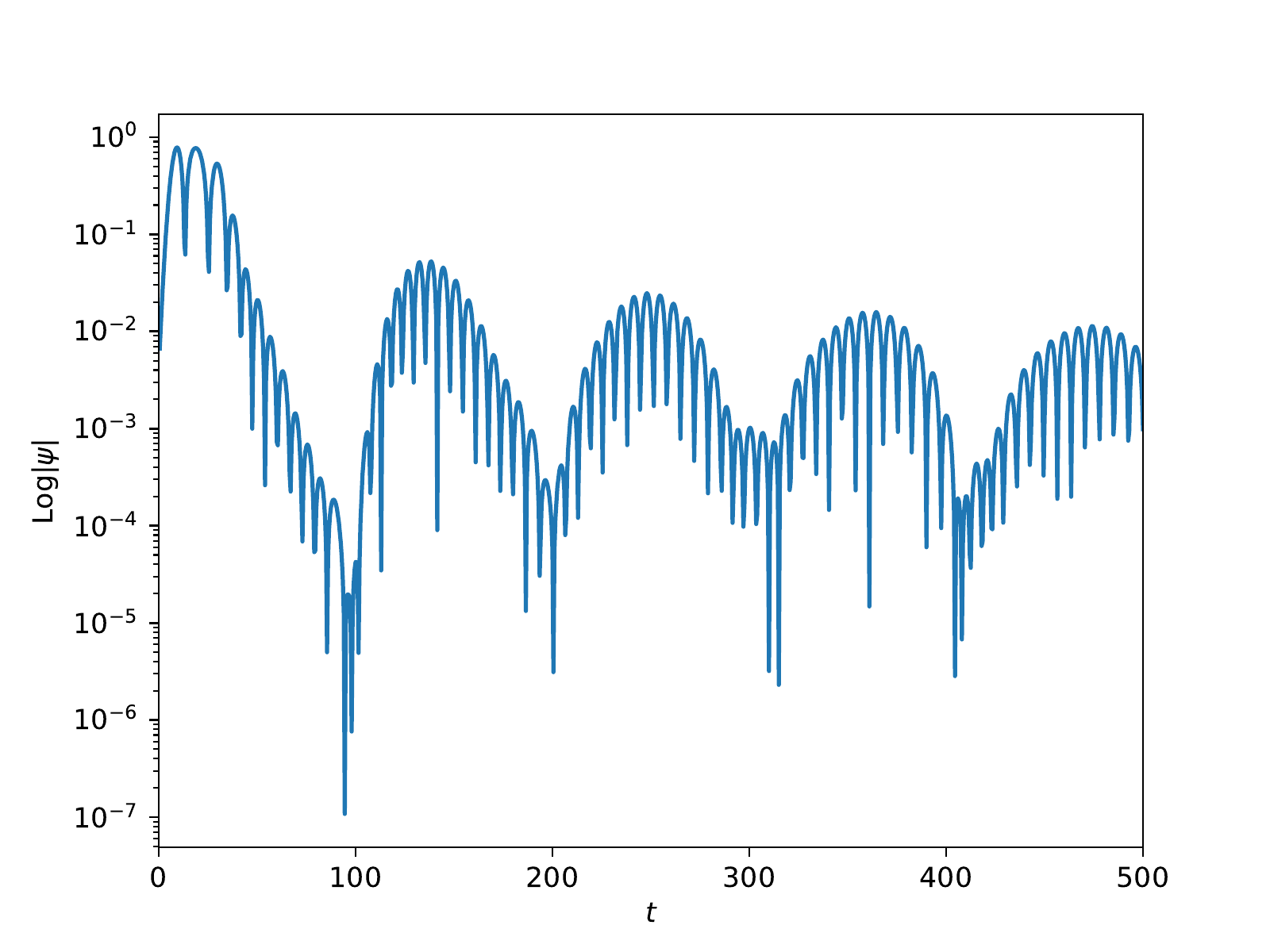}
}
{
\includegraphics[width=0.9\columnwidth]{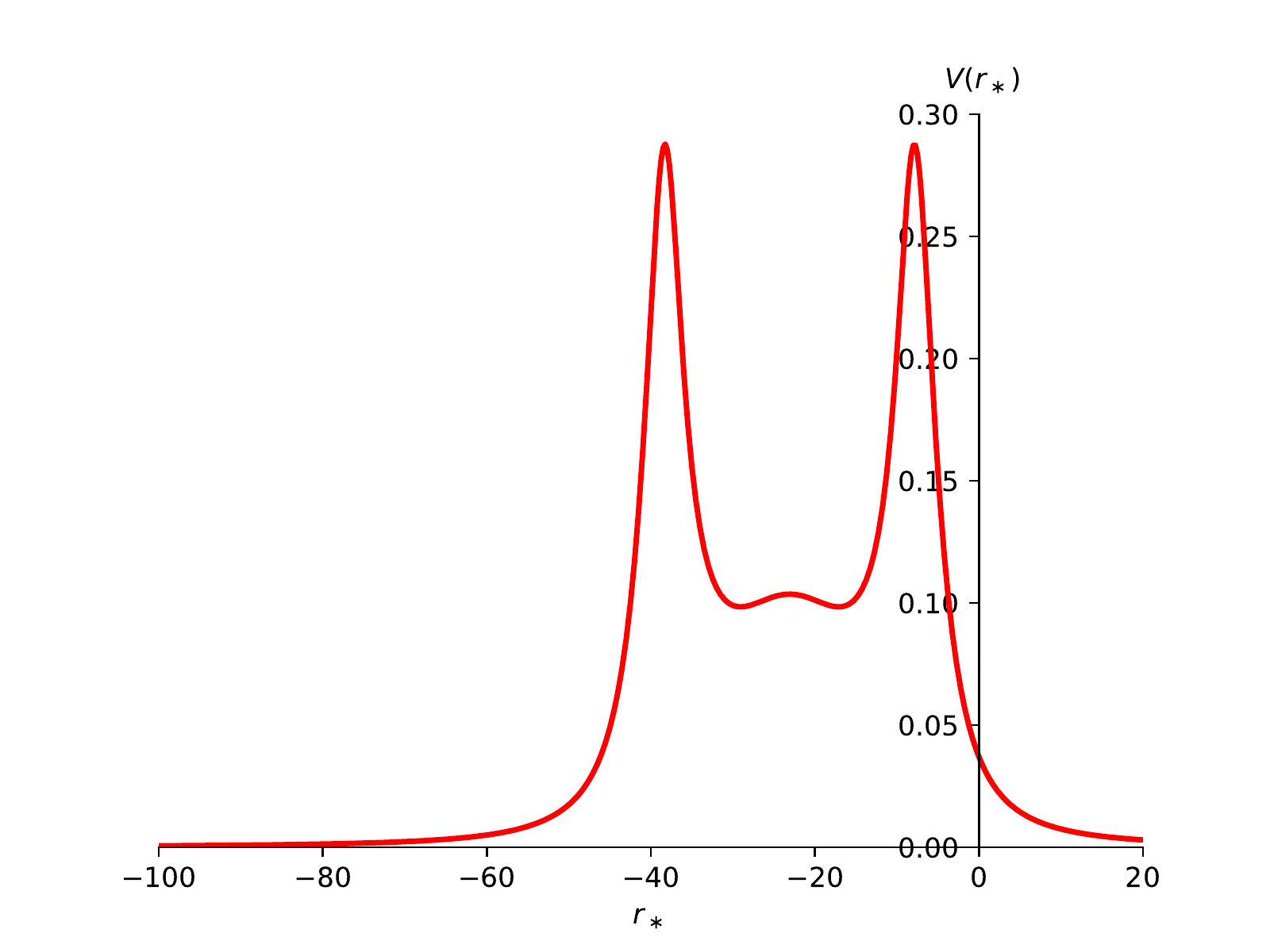}
}
{
\includegraphics[width=0.9\columnwidth]{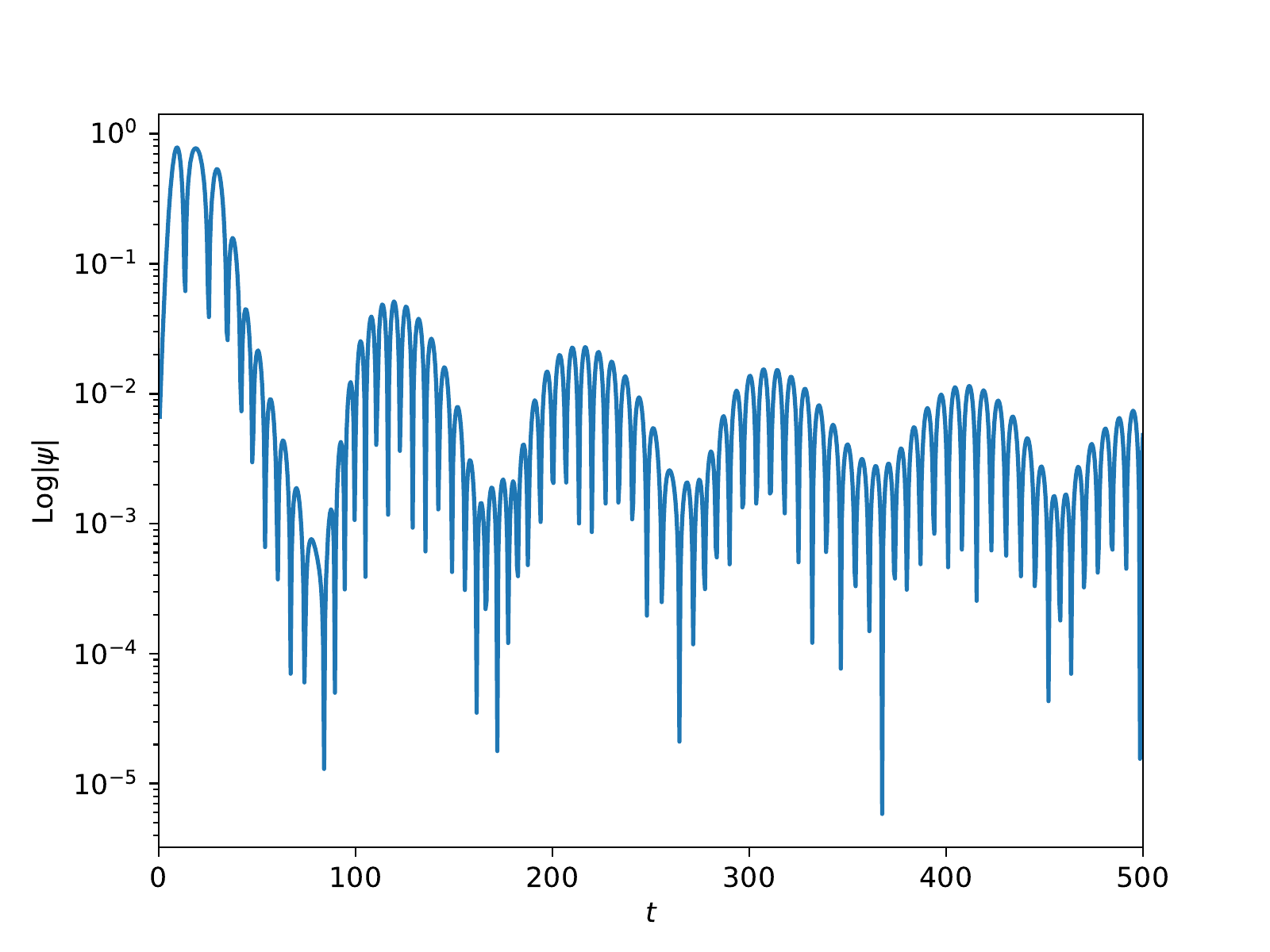}
}
{
\includegraphics[width=0.9\columnwidth]{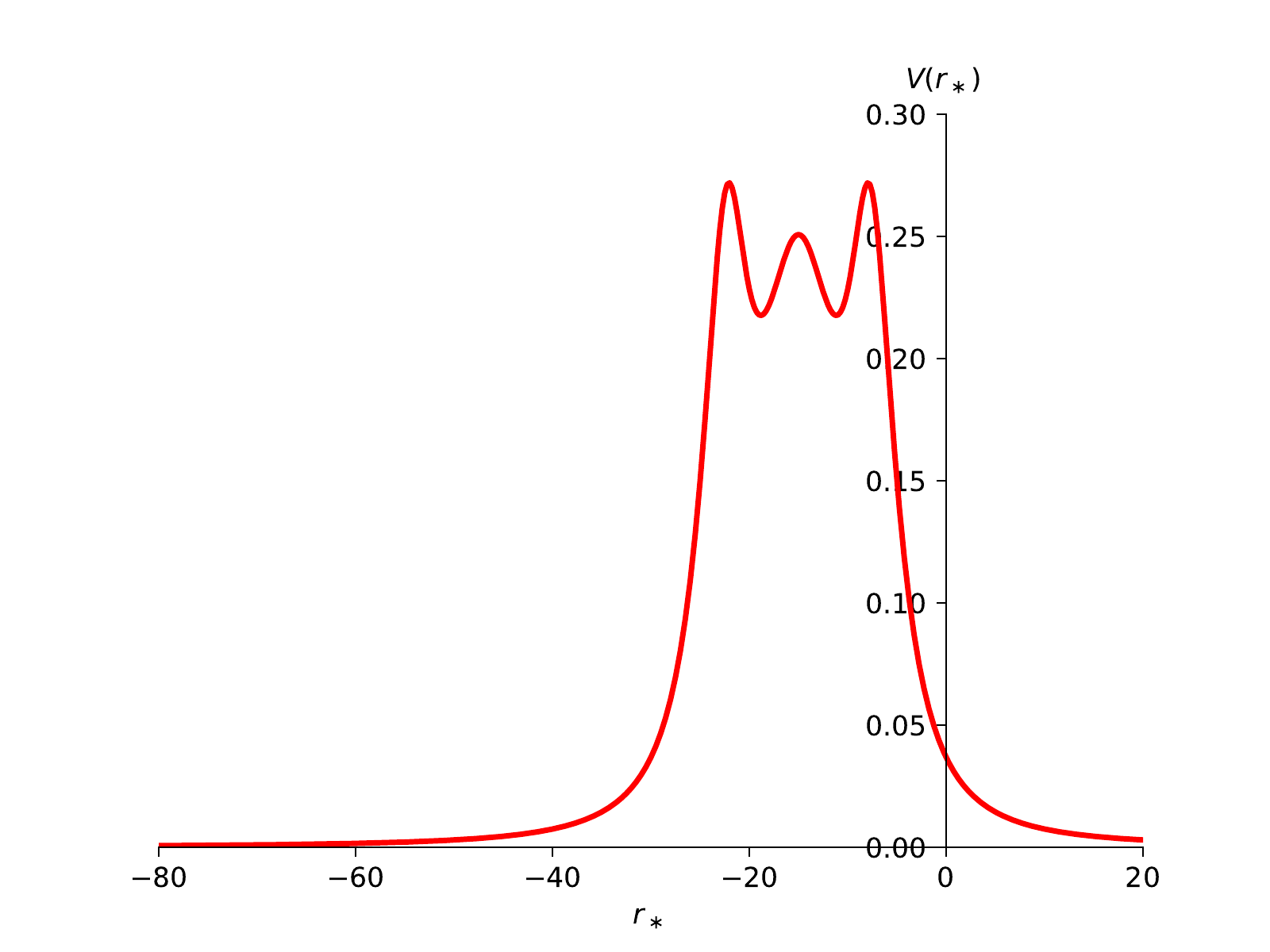}
}
{
\includegraphics[width=0.9\columnwidth]{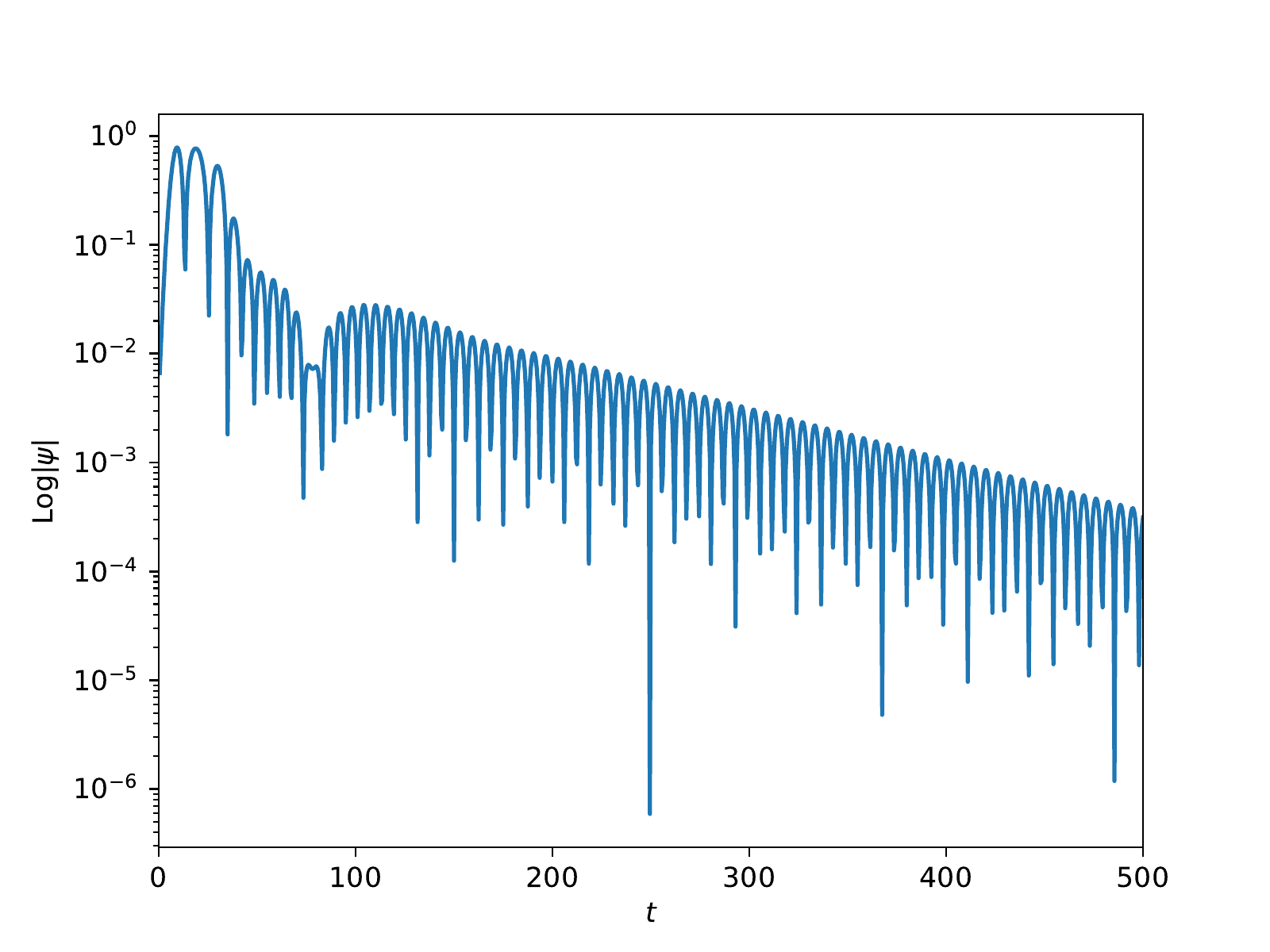}
}
\caption{The effective potential (left panel) and the semi-logarithmic plot of time evolution of scalar perturbations (right panel) with $a =1.02$, $a =1.04$, $a =1.06$, and
$a =1.2$ (from top to bottom) for the two-way traversable wormhole with $M=0.5,l=1$.}
\label{secho}
\end{figure*}

(i) As the parameter $a$ increases, the time interval for the first echo after the initial ringdown becomes shorter, and the time interval for the second echo after the first echo becomes shorter, which shows that the time interval has a strong dependence on the spacetime parameter $a$. This time interval decreases as $a$ increases, so we can predict from this conclusion that when parameter $a$ reaches a certain level, only weak echoes signals can be detected, or even none, because the increase of the parameter $a$ makes the width of the potential well become smaller and smaller, and finally merges into a single-peak potential barrier.

(ii) When the spacetime parameter becomes larger, the amplitudes of the echoes signal only slightly change. This result can be understood by the effective potential function. From Fig. \ref{V101102} and Fig. \ref{V10511}, we can see that as the parameter $a$ increases, the peak value of the potential well only slightly changes. It is because the scalar wave or electromagnetic wave is difficult to escape from the potential well that the echoes are generated. Therefore, potential wells of almost the same height make the probability of scalar waves or electromagnetic waves escaping from the potential well almost the same. This is why the amplitudes of the echoes signal remain almost unchanged.

(iii) Compared with the initial ringdown, the amplitude of the echoes is much smaller. Moreover, the amplitude of the second echo signal is much smaller than the first echo signal, which demonstrates that the amplitude of the echo signal gradually becomes smaller as time progresses. This will make it extremely difficult to find echoes in experiments, and future experiments may require more high-precision experimental instruments to observe it.

Now, let's discuss the echoes characteristics of the scalar field perturbations.
Figure \ref{secho} displays the effective potential of the scalar field and time evolution of scalar field perturbations on the novel black-bounce spacetimes background. One can see that as the parameter $a$ increases, the width and depth of the potential well are decreasing, which is the same as the property of the potential function observed in the electromagnetic field. As we mentioned earlier, three peaks appear during the process of merging the effective potential into one peak for the electromagnetic field. When the parameter $a$ increases to a certain value, the effective potential of the scalar field also appears three peaks. In addition, the peak value of the effective potential of the scalar field is greater than the peak value of the electromagnetic field, which shows that the contribution of the last term of the effective potential under the scalar field perturbation is not negligible for novel black-bounce spacetimes.

In the right column of Fig. \ref{secho}, when the parameters $a=1.02$, $a=1.03$ and $a=1.04$, we see that the most obvious effect is that a clear echo signal appears after the initial quasinormal ringdown. But when $a=1.2$, there is no echo signal after the initial quasinormal ringdown, and there is another type of quasinormal ringdown, which is the quasinormal ringdown of the two-way traversable wormhole. This fact coincides with our prediction when discussing the echo signal generated by the electromagnetic field perturbations. Meanwhile, this behavior is similar to the results of Chowdhury \textit{et al}. \cite{124051}, but they consider the gravitational perturbations in Janis-Newman-Winicour spacetime with naked singularities. As far as the echo signal is concerned, the main feature of the echo signal generated by the scalar field perturbations is almost the same as the echo signal generated by the electromagnetic field perturbations. As the parameter $a$ increases, the time interval between the echoes becomes shorter. Moreover, this time interval is smaller than the corresponding situation in the electromagnetic field. It is worth noting that the amplitude of the echo is similarly smaller than the initial ringdown.

\section{The QNM frequencies of novel black-bounce spacetimes}\label{frequency}
\begin{table*}[t!] \centering
\setlength{\abovecaptionskip}{0.5cm}
\setlength{\belowcaptionskip}{0.2cm}
\caption{The QNM frequencies of electromagnetic field echoes.}
\setlength{\tabcolsep}{3mm}{
\begin{tabular}{ccccc}
\hline\hline
 $l$ \quad \quad &$a$ &Initial ringdown& First echo & Second  echo \\ \hline
\multirow{1}{*}{$l=1$}& 1.02&$0.432511 - 0.139550 i$ & $0.431431 - 0.0909564 i$ & $0.416114 - 0.0801438 i$ \\
 & 1.04& $0.432345 - 0.124975  i$ & $0.420477 - 0.0990151 i$ & $0.375430 - 0.0793338 i$ \\
 & 1.06& $0.426784 - 0.127676  i$ & $0.408921 - 0.0610699 i$ & $0.398697 - 0.0348392  i$ \\
\multirow{1}{*}{$l=2$} &1.02 & $0.750710 - 0.150809 i$ & $0.772872 - 0.1029900  i$ & $0.766563 - 0.0903414 i$\\
 & 1.04& $0.750162 - 0.145696 i$ & $0.766432 - 0.1061910 i$ & $0.762562 - 0.0770380 i$\\
 & 1.06& $0.758609 - 0.141621 i$ & $0.773237 - 0.1025230  i$ & $0.740002 - 0.0941247 i$\\
 \multirow{1}{*}{$l=3$} &1.02 & $1.072990 - 0.189673 i$ & $1.138030 - 0.0924377  i$ & $1.118960 - 0.0707061 i$\\
 & 1.04& $1.065940 - 0.180983 i$ & $1.131770 - 0.0808849 i$ & $1.113860 - 0.0697121 i$\\
 & 1.06& $1.062210 - 0.171786 i$ & $1.123070 - 0.0806182  i$ & $1.105550 - 0.0742054 i$\\
\hline\hline
\end{tabular}}
\vspace{0.4cm}
\label{qnme}
\end{table*}

\begin{table*}[t!] \centering
\setlength{\abovecaptionskip}{0.4cm}
\setlength{\belowcaptionskip}{0.2cm}
\caption{The QNM frequencies of scalar field echoes.}
\setlength{\tabcolsep}{3mm}{
\begin{tabular}{ccccc}
\hline\hline
 $l$ \quad \quad &$a$ &Initial ringdown& First echo & Second  echo \\ \hline
\multirow{1}{*}{$l=1$}& 1.02&$0.502548 - 0.139999 i$ & $0.497635 - 0.0910389 i$ & $0.480468 - 0.0793380 i$ \\
 & 1.04& $0.498359 - 0.137980  i$ & $0.497157 - 0.0919687 i$ & $0.369577 - 0.0812431  i$ \\
 & 1.06& $0.491547 - 0.130677  i$ & $0.393255 - 0.0650390 i$ & $0.405710 - 0.0513467  i$ \\
\multirow{1}{*}{$l=2$} &1.02 & $0.804469 - 0.174604 i$ & $0.816541 - 0.1186340 i$ & $0.809547 - 0.0969228 i$\\
 & 1.04& $0.801596 - 0.169572 i$ & $0.811849 - 0.1098820 i$ & $0.801678 - 0.0976120 i$\\
 & 1.06& $0.798871 - 0.163620  i$ & $0.810072 - 0.1059500 i$ & $0.793749 - 0.1013810 i$\\
 \multirow{1}{*}{$l=3$} &1.02 & $1.116940 - 0.226836 i$ & $1.159890 - 0.0941111 i$ & $1.146080 - 0.0712003 i$\\
 & 1.04& $1.115040 - 0.216005 i$ & $1.156110 - 0.0824485 i$ & $1.142690 - 0.0743523  i$\\
 & 1.06& $1.113020 - 0.206766  i$ & $1.154810 - 0.0774514  i$ & $1.119260 - 0.0697735 i$\\
\hline\hline
\end{tabular}}
\vspace{0.4cm}
\label{qnms}
\end{table*}
In the previous section, we studied the echo signal of novel black-bounce spacetimes, but we only studied the time-domain profiles of the echoes signal. In order to have a further quantitative understanding about the echoes signal, we will study its QNM frequency. On the other hand, when studying the echoes signal of scalar field perturbations, we found that when the spacetime parameter $a$ increases to a certain threshold, the echo signal transforms into the QNM of the wormhole. Therefore, it is also meaningful to quantitatively study the QNM frequency of wormholes, because it directly reflects the unique properties of compact objects; i.e. QNM is the characteristic ``sound'' of compact objects, which has the characteristics of complex frequency, and it can help us further understand the nature of compact objects. The echoes signal are obviously not determined by a single QNM. Therefore, the WKB method \cite{Schutz:1985km,Konoplya2003,hwkb1,hwkb2,hwkb3} cannot be used to calculate the QNM frequency in this case.
We use the Prony method to extract the QNM frequency by the damped exponents \cite{smyth,smyth2,Djermoune,prony}
\begin{equation}
\psi(t) \simeq \sum_{i=1}^{p} C_{i} e^{-i \omega_{i} t}.
\end{equation}

In Table \ref{qnme}, we use the Prony method to calculate the QNM frequencies of the echoes signal in the electromagnetic field perturbations. We list the QNM frequencies of initial ringdown, first echo, and second echo under different angular quantum numbers $l$ and different spacetime parameters $a$. In particular, the result for $l=1$ is extracted from Figs. \ref{echo102}-\ref{echo106}. Since the echoes of $l=2$ and $l=3$ have similar results to those of $l=1$, we have not given their echo pictures, but the analysis of their QNM frequencies is still necessary. From Table \ref{qnme}, we can see that for the case of $l=1$, as the parameter $a$ increases, the real and imaginary parts of the QNM frequencies are both decreasing. In other words, both the oscillation frequency and damping rate are decreasing. Moreover, as we have seen in Figs. \ref{echo102}-\ref{echo106}, initial ringdown has a larger oscillation frequency and damping rate. At the same time, because the second echo contains less energy than the first echo, its oscillation frequency and damping rate are smaller than the first echo. For the case of $l=2$ and $l=3$, the oscillation frequency of the first echo signal is larger than that of the initial ringdown signals, but the damping rate is smaller than that of the initial ringdown signals. The oscillation frequency and damping rate of the second echo signal are still smaller than the first echo. When fixing the parameter $a$, we can see that with the increase of angular quantum numbers $l$, both the real and imaginary parts of the QNM frequencies increase. In Table \ref{qnms}, we list QNM frequencies in the scalar field perturbations. The results presented in Table \ref{qnms} have similar behaviors to the QNM frequencies in the electromagnetic field perturbations.

\begin{figure*}[!htbp]\centering
{
\includegraphics[width=0.9\columnwidth]{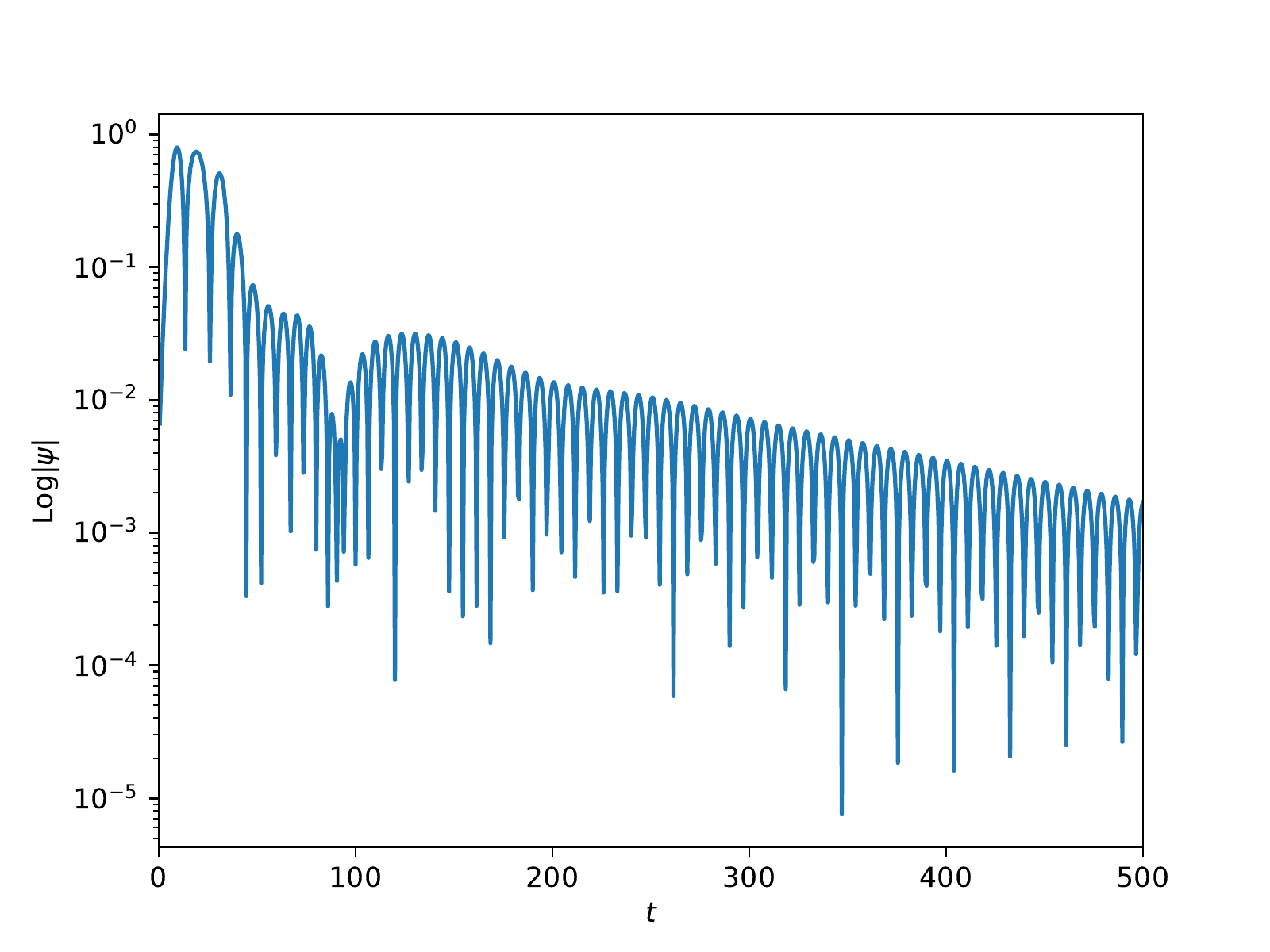}
}
{
\includegraphics[width=0.9\columnwidth]{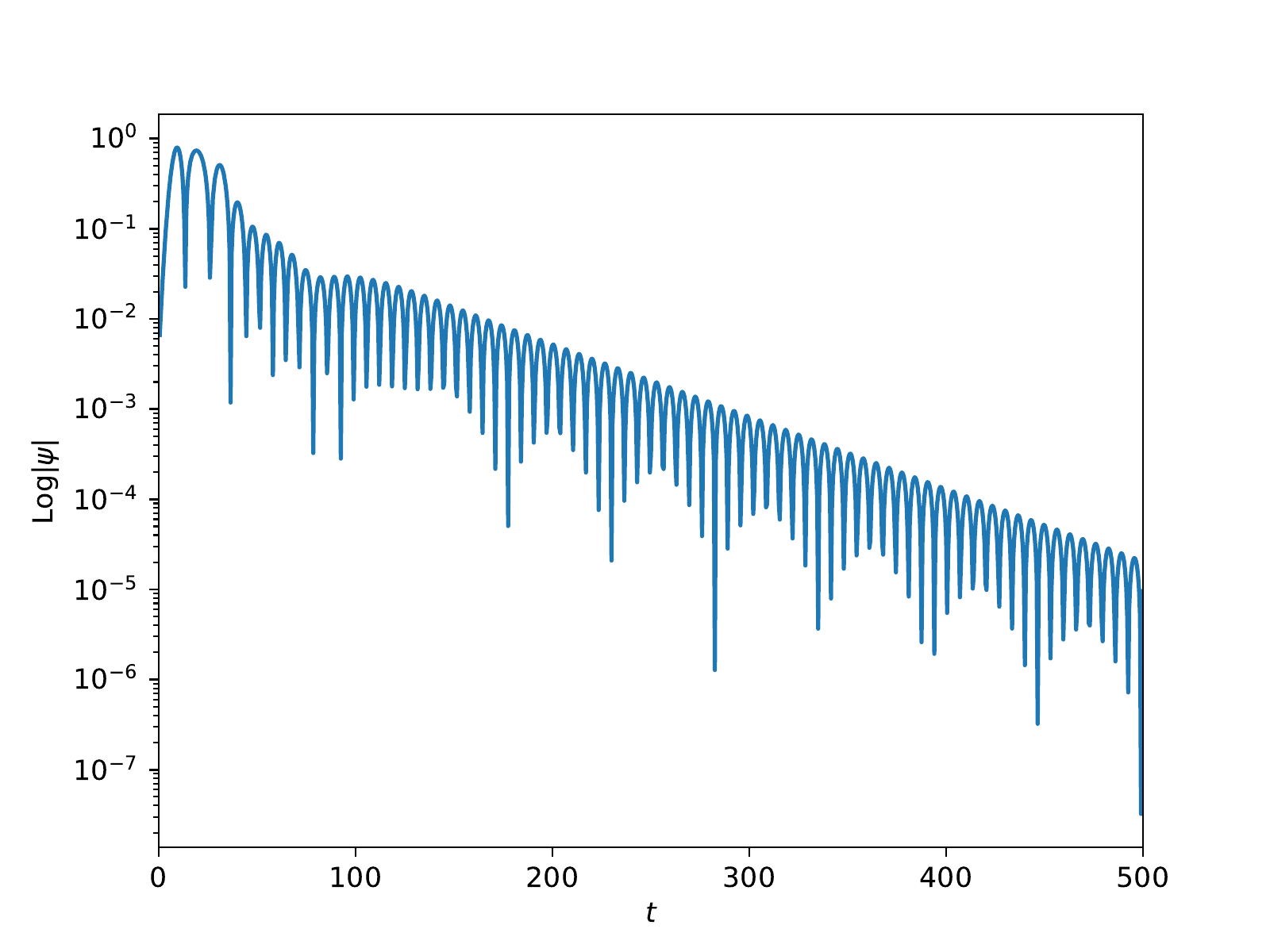}
}
{
\includegraphics[width=0.9\columnwidth]{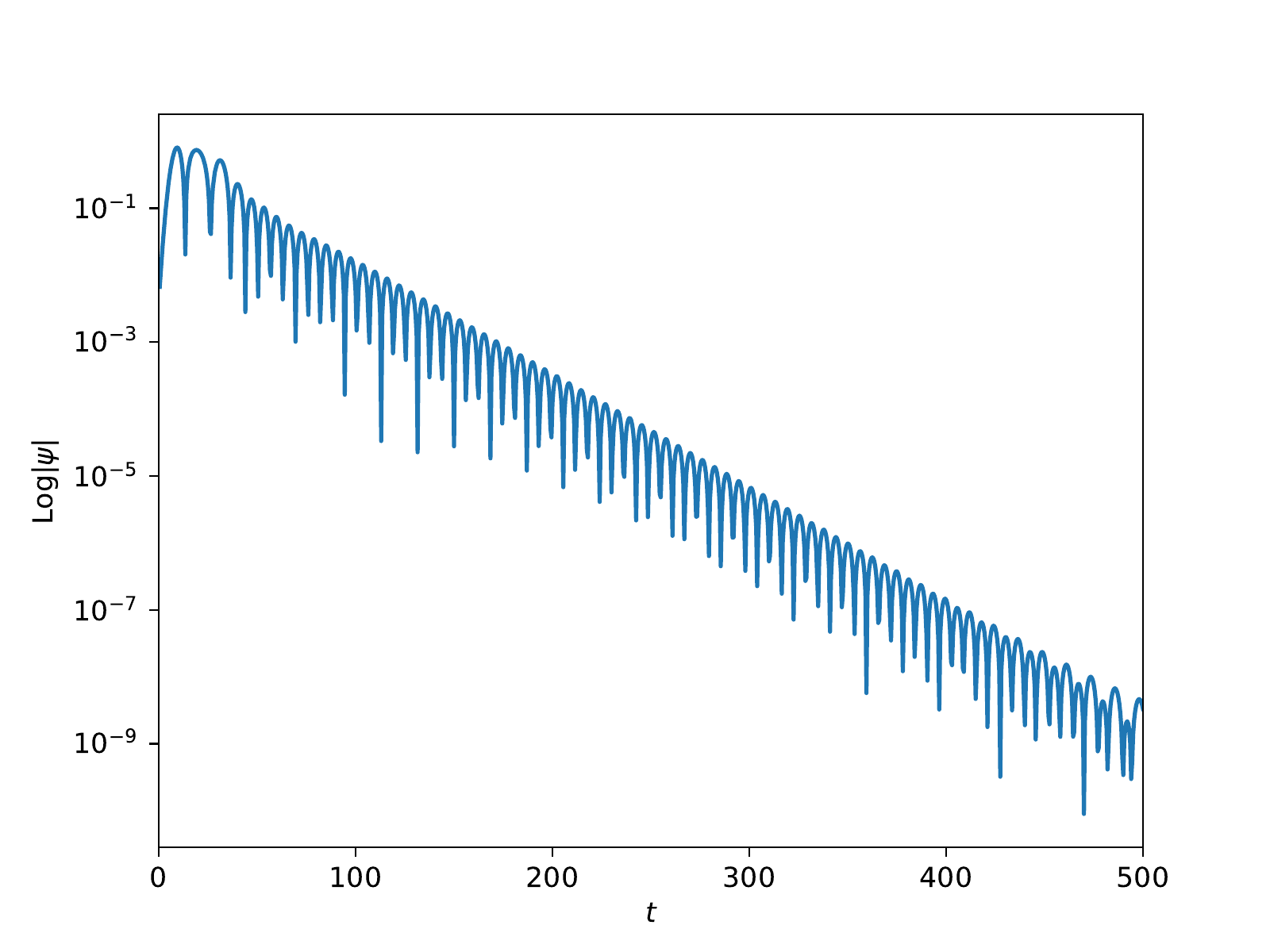}
}
{
\includegraphics[width=0.9\columnwidth]{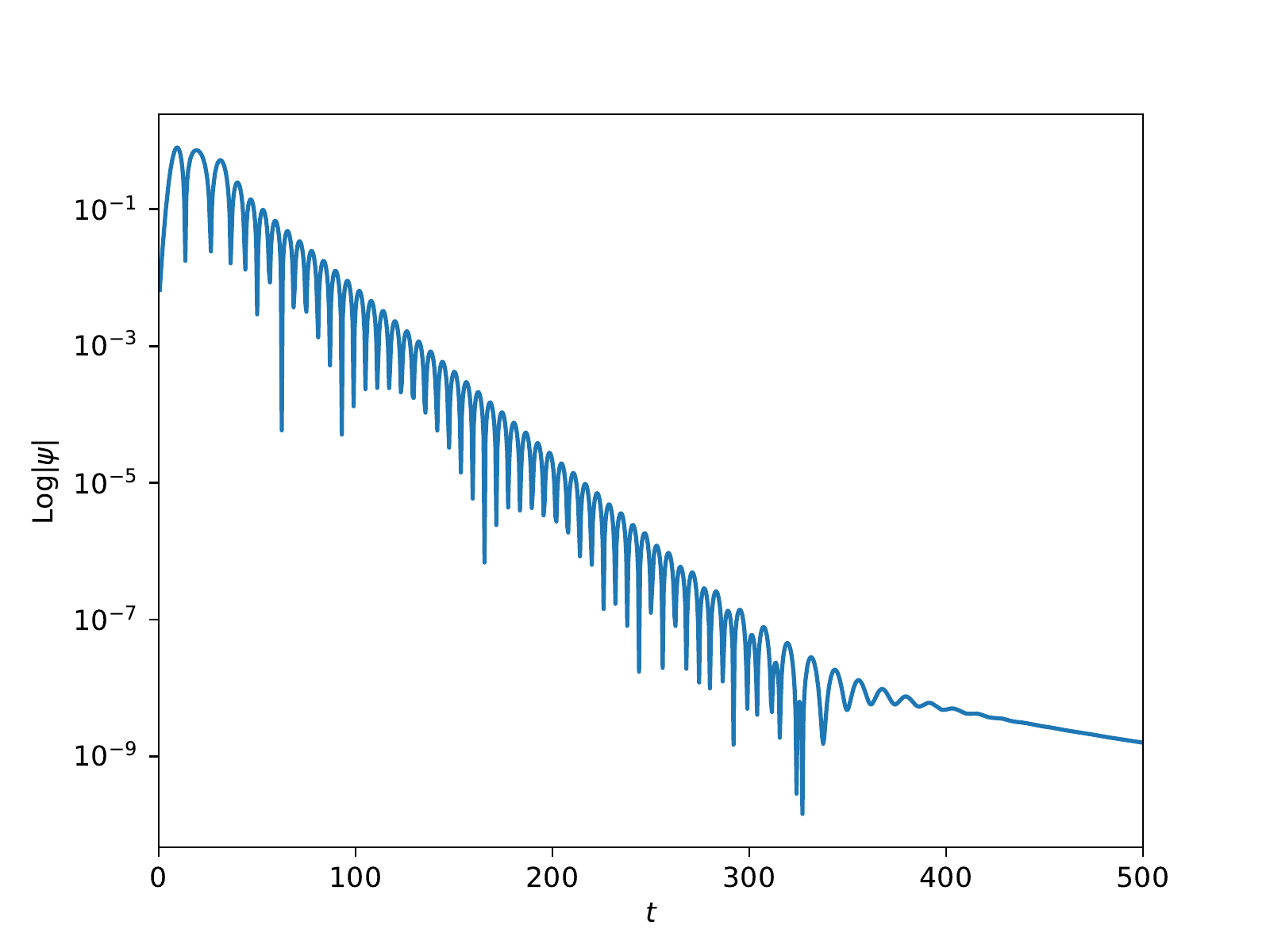}
}
\setlength{\belowcaptionskip}{0.4cm}
\caption{The semi-logarithmic plot of time evolution of electromagnetic field perturbations with $a =1.14$, $a =1.2$ $a =1.3$, and $a =1.4$ (from left to right and from top to bottom) for the two-way traversable wormhole with $M=0.5,l=1$.}
\label{wormqnm}
\end{figure*}

In Fig. \ref{wormqnm}, we present the wormhole's quasinormal ringing with the larger spacetime parameter $a$ for electromagnetic field perturbations. In Fig. \ref{wormqnm}, the most striking feature is that the difference in damping rates is particularly obvious for different $a$. For smaller $a$, perturbation decays are slower, and perturbation decays are faster for larger $a$. In order to quantitatively verify this qualitative behavior, we calculate its QNM frequency. Table \ref{qnmw} lists the wormhole's QNM frequencies under different parameter $a$ and angular quantum numbers $l$. It can be seen from Table \ref{qnmw} that the real and imaginary parts of the QNM frequencies are proportional to the parameter $a$. The characteristics presented by these results are consistent with the results in Fig. \ref{wormqnm} for the same $l$. When the parameter $a$ is fixed and $l$ is increased, the real part of the QNM frequencies are proportional to $l$, but the imaginary part of the QNM frequencies vary irregularly. The ``echoes" in Table \ref{qnmw} mean that there is still the echo signal at this time. From Table \ref{qnmw}, we also find that the threshold of echo disappearing is different for different angular quantum numbers $l$.  When $l$ is smaller, the threshold for the echo to turn into a wormhole QNM is also smaller.

\begin{table*}[t!] \centering
\setlength{\abovecaptionskip}{0.4cm}
\setlength{\belowcaptionskip}{0.2cm}
\caption{The QNM frequencies of electromagnetic field perturbation for traversable wormhole.}
\setlength{\tabcolsep}{3mm}{
\begin{tabular}{ccccc}
\hline\hline
\quad &$a$ &$l=1$& $l=2$ & $l=3$ \\ \hline
 & 1.14& $0.440865 - 0.00720609  i$ & $Echoes $ & $Echoes  $ \\
 & 1.16& $0.455959 - 0.0106964  i$ & $Echoes $ & $Echoes  $ \\
 & 1.18& $0.468484 - 0.0144827  i$ & $0.793189 - 0.00767738 i$ & $Echoes  $ \\
 & 1.20& $0.478865 - 0.0184685  i$ & $0.811127 - 0.0121937 i$ & $1.13828 - 0.0111811  i$\\
 & 1.26& $0.500678 - 0.0308834  i$ & $0.848779 - 0.0267356 i$ & $1.19011 - 0.0439598 i$\\
 & 1.30& $0.509679 - 0.0389546 i$ & $0.862189 - 0.0369372  i$ & $1.21449 - 0.0592475  i$\\
 & 1.40& $0.519821 - 0.0564344  i$ & $0.882812 - 0.0608442  i$ & $1.27385 - 0.1159980 i$\\
\hline\hline
\end{tabular}}
\vspace{0.5cm}
\label{qnmw}
\end{table*}
\section{Summary} \label{sec5}
In summary, we study the quasinormal ringdown of novel black-bounce spacetimes and observe clear echoes signal. By considering the perturbation of electromagnetic field and scalar field, we derived the motion equation and the effective potential. Then by using the finite difference method for numerical calculation, the quasinormal ringdown and echoes are presented. According to our numerical results, the following summary is obtained:

(a) With the increase of the space-time parameter $a$, the effective potential goes through three stages: A single peak corresponds to a regular black hole, double peak corresponds to a two-way wormhole, and single peak corresponds to the larger $a$.

(b) For $0\leq a\leq2M$, the quasinormal ringdown of the novel black-bounce spacetimes is proposed. There is no echo signal, because, whether it is the Schwarzschild black hole, the regular black hole or the one-way wormhole, they all have the horizon.

(c) For $a>2M$, the clear and unique echoes signal appears after the initial ringdown, but because the increase of the parameter $a$ makes the width of the potential well become smaller and smaller, the time interval between echoes gradually decreases, which is contrary to the behavior of the wormholes echo signal given by Ref. \cite{Liu2020qia}. Potential wells having almost the same height cause scalar waves or electromagnetic waves to escape from potential wells with almost the same probability. Therefore, with the change of the parameter $a$, the amplitudes of the echoes signal are only slightly changed, but gradually narrowing and shallower potential well may result in lower QNM frequencies.

(d) When $a$ is much larger than $2M$, the potential well disappears and becomes a single-peak potential barrier. Therefore, the echo signal no longer exists, but it turns into a quasinormal ringdown of the two-way traversable wormhole. Through the study of QNM frequency to wormhole, we find that the QNM frequencies are proportional to the spacetime parameter $a$, and the threshold for this transition is proportional to the angular quantum number $l$. The QNM results of wormholes may be used as a probe for detecting wormholes in the future.

This work is only considered scalar and electromagnetic perturbations, and we find that the echoes signals of two perturbations have very similar characteristics. The gravitational field perturbations may also be very interesting, because these studies \cite{124051,2020jml,2021kwb,Dong:2020odp,Dey:2020lhq,Dey:2020pth,Oshita:2020dox,Chatzifotis:2021pak} have shown that echoes signals can also be generated under gravitational perturbations. The gravitational radiation excited by gravitational perturbations is much stronger than the external field perturbations. Therefore, we believe that if the perturbations of the gravitational field are considered in the novel black-bounce spacetimes, the echoes signal can also be observed.

\begin{acknowledgments}
We are very grateful to R. Moderski and M. Rogatko; R. A. Konoplya and A. Zhidenko; C. Gundlach and R. H. Price; E. H. Djermoune; M. R. Osborne and G. K. Smyth for kindly providing us with useful code. We would also like to thank V. Cardoso; K. A. Bronnikov; C. Vlachos and K. Destounis; A. Chowdhury; Hang Liu and Jian-Pin Wu; Jun-Jin Peng for helpful correspondence.
We earnestly thank the reviewer for the thorough reading of our manuscript and for constructive comments. This research was funded by the National Natural Science Foundation of China (Grant Nos. 11465006 and 11565009) and the Natural Science Special Research Foundation of Guizhou University (Grant No. X2020068).
\end{acknowledgments}
\nocite{*}

\end{document}